\documentclass{aa}

\usepackage{eurosym,rotating, booktabs}
\usepackage{graphicx}
\usepackage{amsmath}
\usepackage{txfonts}
\usepackage{csquotes}
\usepackage{hyperref}
\usepackage{xurl}
\usepackage{longtable}
\usepackage{threeparttable}
\usepackage{comment}

\begin{document}

\title{FAUST XX. The chemical structure and temperature profile\\ of the IRAS 4A2 hot corino at 20-50 au}

\author{J. Frediani
          \inst{1,2,3} 
        \and
        M. De Simone
        \inst{2,4} 
        \and
        L. Testi
          \inst{1,4}
        \and
        L. Podio
        \inst{4}
        \and
        C. Codella 
        \inst{4}
        \and
        C. J. Chandler
        \inst{5}
        \and
        C. Ceccarelli
          \inst{6}
        \and
        L. Loinard
        \inst{7,8,9}
        \and
        A. L\'opez-Sepulcre
        \inst{6,10}
        \and
        B. Svoboda
        \inst{5}
        \and
        N. Sakai
        \inst{11}
        \and
        L. Chahine
        \inst{6}
        \and
        Y. Aikawa
        \inst{12}
        \and 
        E. Bianchi
        \inst{4}
        \and
        M. Bouvier
        \inst{13}
        \and
        L. Cacciapuoti
        \inst{2}
        \and
        P. Caselli
        \inst{14}
        \and
        S. B. Charnley
        \inst{15}
        \and
        I. Jimenez-Serra
        \inst{16}
        \and
        D. Johnstone
        \inst{17,18}
        \and
        G. Sabatini
        \inst{4}
        \and
        Y. Shirley
        \inst{19}
        \and
        S. Yamamoto
        \inst{20}
}  
   \institute{DIFA, Dipartimento di Fisica e Astronomia, Università degli Studi di Bologna, Via Gobetti 93/2, I-40129 Bologna, Italy \\
   \email{jenny.frediani@astro.su.se}
   \and
   ESO, Karl Schwarzchild Str. 2, 85748 Garching bei München, Germany\\
   \email{marta.desimone@eso.org}
   \and
   Department of Astronomy, Stockholm University, AlbaNova University Centre, 106 91 Stockholm, Sweden
   \and
   INAF, Osservatorio Astrofisico di Arcetri, Largo E. Fermi 5, 50125 Firenze, Italy
   \and
    National Radio Astronomy Observatory, 1011 Lopezville Rd, Socorro, NM 87801, USA
    \and
    Univ. Grenoble Alpes, CNRS, IPAG, 38000 Grenoble, France
    \and 
    Instituto de Radioastronom\'{i}a y Astrof\'{i}sica, Universidad Nacional Aut\'{o}noma de M\'{e}xico, A.P. 3-72 (Xangari), 8701, Morelia, Mexico 
    \and
     Black Hole Initiative at Harvard University, 20 Garden Street, Cambridge, MA 02138, USA
    \and
    David Rockefeller Center for Latin American Studies, Harvard University, 1730 Cambridge Street, Cambridge, MA 02138, USA
    \and
    Institut de Radioastronomie Millim\'etrique (IRAM), 300 rue de la Piscine, 38406 Saint-Martin-d'H\`eres, France
    \and
    The Institute of Physical and Chemical Research (RIKEN), 2-1, Hirosawa, Wako-shi, Saitama 351-0198, Japan
    \and
    Department of Astronomy, The University of Tokyo, Bunkyo-ku, Tokyo 113-0033, Japan
    \and
    Leiden Observatory, Leiden University, P.O. Box 9513, 23000 RA Leiden, The Netherlands
    \and
    Center for Astrochemical Studies, Max-Planck-Institut für Extraterrestrische Physik, Gießenbachstraße 1, 85748 Garching, Germany
    \and
    Astrochemistry Laboratory, Code 691, NASA Goddard Space Flight Center, 8800 Greenbelt Road, Greenbelt, MD 20771, USA
    \and
    Centro de Astrobiología (CAB), INTA-CSIC, Carretera de Ajalvir km 4, Torrejón de Ardoz, 28850 Madrid, Spain
    \and
    NRC Herzberg Astronomy and Astrophysics, 5071 West Saanich Road, Victoria, BC, V9E 2E7, Canada
    \and 
    Department of Physics and Astronomy, University of Victoria, Vic- toria, BC, V8P 5C2, Canada
    \and 
    Steward Observatory, 933 N Cherry Ave., Tucson, AZ 85721 USA
    \and
    The Graduate University for Advanced Studies (SOKENDAI), Shonan Village, Hayama,Kanagawa 240-0193, Japan}
%

\titlerunning{FAUST XX. The IRAS 4A2 hot corino at 20-50 au}
\authorrunning{Frediani, De Simone et al.}



\abstract
{Young low-mass protostars often possess hot corinos, compact, hot and dense regions bright in interstellar Complex Organic Molecules (iCOMs). Besides of their prebiotic role, iCOMs can be used as a powerful tool to characterize the chemical and physical properties of hot corinos.}
{Using ALMA/FAUST data we aim to explore the iCOMs emission at $<$ 50 au scale around the Class 0 prototypical hot corino IRAS 4A2.}
{We imaged IRAS 4A2 in six abundant, common iCOMs (CH$_3$OH, HCOOCH$_3$, CH$_3$CHO, CH$_3$CH$_2$OH, CH$_2$OHCHO, and NH$_2$CHO), and derived their emitting size. The column density and gas temperature for each species were derived at 1$\sigma$ from a multi-line analysis by applying a non-LTE approach for CH$_3$OH, and LTE population or rotational diagram analysis for the other iCOMs. Thanks to the unique estimates of the absorption from foreground millimeter dust toward IRAS 4A2, we derived for the first time unbiased gas temperatures and column densities.} 
{We resolved the IRAS 4A2 hot corino finding evidence for a chemical spatial distribution in the inner 50 au, with the outer emitting radius increasing from $\sim$ 22-23 au for NH$_2$CHO and CH$_2$OHCHO, followed by CH$_3$CH$_2$OH ($\sim$ 27 au), CH$_3$CHO ($\sim$ 28 au),  HCOOCH$_3$ ($\sim$ 36 au), and out to $\sim$ 40 au for CH$_3$OH.
Combining our estimate of the gas temperature probed by each iCOM with their beam-deconvolved emission sizes, we inferred the gas temperature profile of the hot corino on scales of 20-50 au in radius, finding a power-law index $q$ of approximately --1.}
{We observed, for the first time, a chemical segregation in iCOMs of the IRAS 4A2 hot corino, and derived the gas temperature profile of its inner envelope.
The derived profile is steeper than when considering a simple spherical collapsing and optically-thin envelope, hinting at a partially optically-thick envelope or a gravitationally unstable disk-like structure.}



\maketitle

\begin{figure*}[htp!]
\centering
\includegraphics[width=\textwidth]{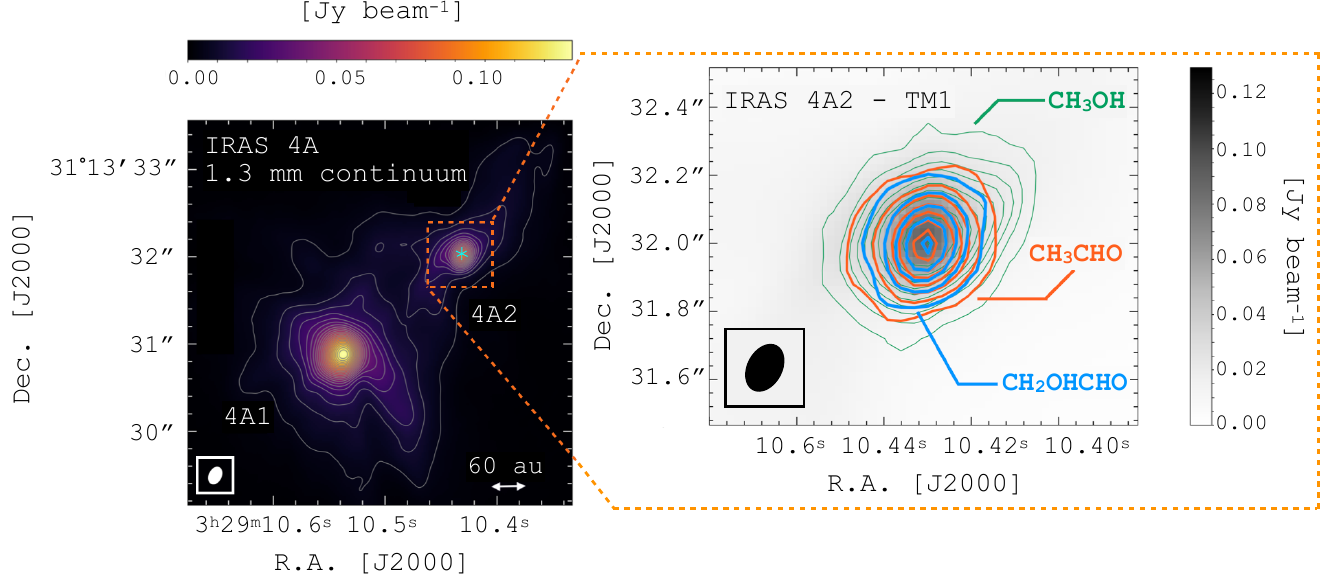}
\caption{Dust continuum and emission of selected iCOMs transitions towards IRAS 4A2. 
\textit{Left}: 1.3 mm dust continuum emission in colour scale and grey contours. First contours and
steps are 3$\sigma$ ($\sigma$ = 0.6 mJy beam$^{-1}$). The cyan star marks the IRAS 4A2 dust peak, 3$^h$29$^m$10$^s$.431
(R.A.) and +31$^{\circ}$13$'$32$\farcs00$ (Dec.). The beam is $0\farcs20\times0\farcs14$ (-26$^{\circ}$). \textit{Right}: Dust continuum (gray scale) overlaid with moment 0 contours maps of CH$_3$OH 16$_{(-2,15)}$-15$_{(-3,13)}$ E (\textit{green}), integrated over +3 to +10 km s$^{-1}$, CH$_3$CHO 13$_{(0,13)}$-12$_{(0,12)}$ E [$v_t$=2] (\textit{orange}), integrated over +4 to +9 km s$^{-1}$, and CH$_2$OHCHO 34$_{(5,30)}$-34$_{(4,31)}$ (\textit{blue}), integrated over +6 to +8 km s$^{-1}$. First contours and steps are 3 and 6$\sigma$ respectively, with $\sigma$ = 3.7 mJy beam$^{-1}$ km s$^{-1}$ for CH$_3$OH, $\sigma$ = 2.7 mJy beam$^{-1}$ km s$^{-1}$ for CH$_3$CHO, and $\sigma$ = 2.3 mJy beam$^{-1}$ km s$^{-1}$ for CH$_2$OHCHO. The beam is $0\farcs17\times0\farcs12$ (-28$^{\circ}$).}
\label{fig:1}
\end{figure*}
\section{Introduction} \label{sec:introduction}
Observations of Class II protostars, $\sim$ Myr old stars with disks, and cosmochemical evidence from our Solar System suggest a rapid evolution of solids into planetary cores, followed by planet formation and disk-planet interaction earlier than originally thought \citep{Johansen2014, Manara2018, Bernabo2022}. While deriving accurate properties of younger Class 0/I disks \citep[$\sim$ 10$^{4-5}$ yr;][]{lada_1987, andre_2000, andre_2002} is prone to very large uncertainties \citep{tung_2024}, numerical simulations suggest that they may harbour already the conditions required for planet formation \citep{lebreuilly_protoplanetary_2021,lebreuilly_synthetic_2024}. Some observations, albeit affected by large uncertainties, seem to support such a claim \citep{sheehan_2018, tychoniec_2020}. 
Therefore, a chemical characterization of the early Class 0/I stages is crucial to understand what a forming planet can inherit 
\citep{caselli_2012, oberg_2021, ceccarelli_ppvii}.

Solar-type Class 0/I sources often possess compact ($<$ 100 au), hot (T $>$ 100 K), and dense (n$_{\rm H}$ $>$ 10$^{7}$ cm$^{-3}$) regions, named \enquote{hot corinos} \citep{ceccarelli_2004}. 
These sources show high gas-phase abundances of interstellar Complex Organic Molecules \citep[iCOMs, saturated C-bearing molecules with at least six atoms and including heteroatoms, such as N and O;][]{vandishoeck_2009, ceccarelli_2017}, liberated through the sublimation of the dust icy mantles \citep{ceccarelli_2023}.
The work by \citet{maury_2014} was pivotal in showing with interferometric observations the hot corino nature of IRAS 2A, where various iCOMs sizes were estimated to be much more compact than the beam size. In fact, due to their compactness, only four hot corinos have been spatially resolved so far: SVS13-A \citep{bianchi_2022}, HH212 \citep{lee_2022}, IRAS 16293-2422 A \citep{maureira_2022}, and B335 \citep{okoda_2022}. In these sources, the iCOMs are either spatially segregated within the resolved structure, or associated with accretion shocks/hot spots. However, for only a few of these chemical species a good physical characterization has been performed. 

IRAS 4A2, the second discovered hot corino source, has an estimated size of about 70 au \citep[from previous unresolved observations;][]{bottinelli_2004a, taquet_2015, lopez_2017, marta_2017, marta_2020}. It is located in the nearby Perseus/NGC 1333 star forming region \citep[$\sim$300 pc;][]{zucker_2018, ortiz_2018}, and is part of a binary system together with IRAS 4A1, ($1\farcs8$, or $\sim$540 au, away), with a total bolometric luminosity of 9.1 L$_{\odot}$ \citep{kristensen_2012, karska_2013}. 
The system exhibits extended (4000 au) molecular outflow cavities, and evidence for a disk wind at 100 au scale \citep{marta_2020b, marta_2024,chahine_2024}.

With this work, we use the iCOMs emission to characterize the inner 50 au of the IRAS 4A2 hot corino as part of the ALMA (Atacama Large sub-Millimeter Array)\footnote{\url{https://www.almaobservatory.org/en/home/}} Large Program (LP) FAUST \citep[Fifty AU STudy of the chemistry in the disc/envelope system of solar-like protostars;][]{codella_2021}. 

The paper is organized as follows. In Sect. \ref{sec:observations} are described the ALMA/FAUST observations, and the line identification of a sample of iCOMs detected toward IRAS 4A2. Section \ref{sec:results} presents our results. We first resolve the iCOMs molecular emission in the hot corino, and retrieve the spatial distribution of the different species (Sect. \ref{subsec:3.1}). Then we derive gas temperatures and column densities with both LTE and non-LTE methods (Sect. \ref{subsec:3.2}). In Sect. \ref{sec:discussion} we discuss the impact of dust continuum emission on the fitted quantities (\ref{subsec:4.1}), and then directly derive a gas temperature profile at 50 au scales (\ref{Sec:4.2}). The conclusions are wrapped up in Sect. \ref{sec:conclusions}. In the appendix can be found, in order of reference in the text, the individual line spectra (Fig. \ref{fig:6} to Fig. \ref{fig:11}), their spectral parameters and fit results (Tab. \ref{tab2}), the results from image plane fitting of chosen lines (Tab. \ref{tab3}), as well as complementary figures (Fig. \ref{fig:12} and Fig. \ref{fig:13}). There are also dedicated appendicies to the adopted methodology for the retrieval of physical parameters from the iCOMs lines (App. \ref{appendixC} and App. \ref{appendixD}), as well as supplementary material for the results and the discussion  (App. \ref{appendixE}).
\section{Observations and line identification}\label{sec:observations}
The observations of IRAS 4A we present here are part of the ALMA Large Program FAUST (PI. S. Yamamoto, 2018.1.01205.L) performed between October 2018 and September 2019 with baselines for the 12-m array from 15.1m to 3.6 km. Bandpass, flux, and phase calibrators are J0237+2848, J0336+3218 and J0328+3139, respectively. 
The map phase center is at R.A. (J2000) = 03${^h}$29$^{m}$10$^{s}$.539, and Dec. (J2000) = +31$^{\circ}$13'30$\farcs92$.
We used the wide-band spectral windows at 230 GHz (Setup 1, hereafter S1), and at 240 GHz (Setup 2, hereafter S2), both with 1875 MHz bandwidth and 1.1 MHz ($\sim$1.4 km s$^{-1}$) of spectral resolution \citep{codella_2021}. 
The data were calibrated using the ALMA calibration pipeline in the Common Astronomy Software Applications package (\texttt{CASA})\footnote{\url{https://casa.nrao.edu}}, with an additional calibration routine to correct for the T$_{sys}$ normalization issue\footnote{\url{https://help.almascience.org/kb/articles/what-errors-could-originate-from-the-correlator-spectral-normalization-and-tsys-calibration}}.
Phase and amplitude self-calibration were performed on the continuum, generated using manually detected line-free continuum channels and applied to the cube (Chandler et al. in prep.).
We cleaned and imaged the continuum-subtracted line cubes with \texttt{CASA} (V6.5.6) using a \texttt{briggs} weighting (\texttt{robust} = 0.5), multiscale deconvolution (scales = [0, 5, 15, 30, 60]), and \texttt{automasking}. The resulting synthesized beams are $0\farcs21\times0\farcs14$ (PA = -3$^{\circ}$) for S1, and $0\farcs17\times 0\farcs12$ (PA = -28$^{\circ}$) for S2.  We then primary beam corrected the cubes. The 7m ACA data are available only for S2, where we estimated a flux loss of about 25\% over 0$\farcs6$. 
For consistency, we proceeded analyzing the 12m configuration data alone for both setups. The absolute flux error is $\sim$20\%, which includes the calibration uncertainty and an additional error for the spectral baseline determination.

We extracted the spectrum obtained using each of the two setups in the IRAS 4A2 dust continuum emission peak position: R.A. (J2000) = 3${^h}$29$^{m}$10$^{s}$.431 and Dec. (J2000) +31$^{\circ}$13'32$\farcs00$. (Fig. \ref{fig:1}). We searched for the most abundant iCOMs, using the Cube Analysis and Rendering Tool for Astronomy package (\texttt{CARTA}; V4.0.0)\footnote{\url{https://cartavis.org}},  
namely methanol (CH$_3$OH), methyl formate (HCOOCH$_3$, or CH$_3$OCHO; hereafter we will use the former nomenclature), 
acetaldehyde (CH$_3$CHO), formamide (NH$_2$CHO), ethanol (CH$_3$CH$_2$OH, or C$_2$H$_5$OH; afterwards we will use the former nomenclature), 
and glycolaldehyde (CH$_2$OHCHO). 
We identified the lines manually using the CDMS \citep{muller_2005} and JPL \citep{pickett_1998} catalogs, verifying that all predicted transitions of the queried molecule were detected. We consider a detection if the signal-to-noise ratio is above 5. 
In total we detected 74 emission lines (see Tab. \ref{tab2}): 9 of CH$_3$OH (E$_{\rm u}$=61-537 K), 24 of HCOOCH$_3$ (E$_{\rm u}$=114-354 K), 5 of CH$_3$CHO (E$_{\rm u}$=96-487 K), 19 of CH$_3$CH$_2$OH (E$_{ u}$=78-437 K), 6 of NH$_2$CHO (E$_{\rm u}$=78-115 K), and 11 of CH$_2$OHCHO (E$_{\rm u}$=242-489 K), where E$_{\rm u}$ is the transition upper-state energy.
\section{Results} \label{sec:results}

\subsection{Spatial segregation of iCOMs}\label{subsec:3.1}
To study the spatial distribution of the identified iCOMs, we produced and compared integrated intensity maps of different species using isolated transitions with similar E$_{\rm u}$, so as to avoid excitation biases.
Figure \ref{fig:1} shows, on the left, the IRAS 4A system in 1.3 mm dust continuum emission, and, on the right, a zoom-in of the same map onto IRAS 4A2, with over-plotted colored contours of the integrated intensity maps of three selected iCOMs transitions: CH$_3$OH 16$_{(-2,15)}$-15$_{(-3,13)}$ E (E$_{\rm u}$=338 K); CH$_3$CHO 13$_{(0,13)}$-12$_{(0,12)}$ E (E$_{\rm u}$=461 K); CH$_2$OHCHO 34$_{(5,30)}$-34$_{(4,31)}$  (E$_{\rm u}$=344 K). The moment 0 maps of all the isolated transitions for each species are in Fig. \ref{fig:13}. 
These maps show that the various iCOMs trace different scales.
To quantify this, we estimated the emitting radius associated with each species by performing a 2D Gaussian fit of the emission on the integrated intensity maps using the \texttt{imfit} task in \texttt{CASA}. 
For the few transitions whose emission is clearly associated with IRAS 4A2 only, avoiding any contamination from the companion IRAS 4A1, we also performed a Gaussian fit on the visibility plan with the \texttt{uvmodelfit} task in CASA. 
Since the fit results in the visibility plane are consistent within $\pm$ 1--2 au with the image plane determinations, we analyzed all the isolated lines using only the image plane results (Tab. \ref{tab3}).
Some examples of the fitted images and of the residual maps are in Fig. \ref{fig:12}. For all transitions, the residual maps of the fit show no emission above 2-3$\sigma$. 
From the beam-deconvolved major ($\theta_{\rm M}$) and minor ($\theta_{\rm m}$) diameters of the fitted Gaussian ellipses (Tab. \ref{tab3}), we derived a radius of the emitting line as half of the geometric average of $\theta_{\rm M}$ and $\theta_{\rm m}$. 

\begin{figure}[t]
\centering
\includegraphics[width=0.9\columnwidth]{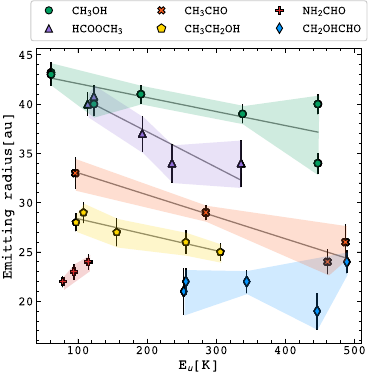}
\caption{iCOMs beam-deconvolved radius as a function of the line upper-state energy. The coloured markers correspond to a chosen sample (see Sect. \ref{subsec:3.1} and Fig. \ref{fig:13}) of imaged and fitted iCOMs transitions. The grey solid lines indicate the linear fit performed on the derived radius. The shaded colored regions highlight the emitting regions associated with each iCOM. 
}\label{fig:2}
\end{figure}

Figure \ref{fig:2} shows the beam-deconvolved radius (in au) for each isolated transition versus the line E$_{\rm u}$. The iCOMs display different emitting sizes, in other words, the observed transitions must trace a different outer radius in the hot corino, where the line optical depth is equal to or larger than 1.
Therefore, the estimated emitting radii within error bars (shaded regions in Fig. \ref{fig:2}) indicate that we are probing different regions of the hot corino of IRAS 4A2, extending out to an outer radius which spans between 16 and 40 au, depending on the considered species. We note, however, that the exact geometry of this molecular emission cannot be unambiguously determined with a 2D Gaussian fit. 
A higher angular resolution coupled with a detailed chemo-physical model of the molecular line emission are needed to resolve this ambiguity.
To estimate the dependence of the emitting size versus the transition, we linearly fitted the derived radius with respect to E$_{\rm u}$, except for CH$_2$OHCHO and NH$_2$CHO due to their partially unresolved emission. 
The fit shows a small variation of the size as a function of E$_{\rm u}$, with negative slopes deviating from zero by factors 0.014 $\pm$ 0.003 (CH$_3$OH, $\tilde{\chi}^2$ = 4.7), 0.04 $\pm$ 0.01 (HCOOCH$_3$, $\tilde{\chi}^2$ = 1.0), 0.010 $\pm$ 0.0022 $\pm$ 0.005 (CH$_3$CHO, $\tilde{\chi}^2$ = 1.4), and 0.016 $\pm$ 0.005 (CH$_3$CH$_2$OH; $\tilde{\chi}^2$ = 0.4). 
We therefore computed a weighted average emission region radius per species ($\bar{r}$; Tab. \ref{tab1}) as a representative emitting size. 

Another possibility is that we are observing different molecular abundances at limited sensitivity. This means that we are able to detect the less abundant species only at the largest densities, hence closer to the protostar at small emitting radii. Among the various detected species, methanol (CH$_3$OH) is known to be the most abundant iCOM, and tracer of the protostellar environment at different scales (e.g. also outflows). Acetaldehyde (CH$_3$CHO), formamide (NH$_2$CHO), and glycolaldehyde (CH$_2$OHCHO), instead, seem to share similar lower abundances, and to trace the hot corino region only \citep[see e.g.][]{taquet_2015, lopez_2017, belloche_2020}. While we cannot exclude a priori the aforesaid possibility, the fact that we observe a difference in the emitting size for all the iCOMs, also among the compact species, strongly suggests that the observed chemical segregation in IRAS 4A2 is real. On top of that, it should be noted that the observed difference in emitting sizes holds independently of the upper-state energy E$_{\rm u}$, as we compared different iCOMs transitions at similar E$_{\rm u}$ precisely to account for gas excitation conditions.

A similar stratification has been observed in other two hot corinos, B335 \citep{okoda_2022}, between CH$_3$OH and NH$_2$CHO, and HH212, with CH$_3$OH, NH$_2$CHO, and CH$_3$CHO, \citep{lee_2022}, as well as toward some \enquote{hot cores} in high massive star forming regions \citep{Jim_nez_Serra_2012, calcutt_2014, Gieser_2019,  Gieser_2021}. \citet{bianchi_2022} and \citet{lee_2022} associated this segregation to the different binding energy of the species. To prove this in IRAS 4A2, binding energies must be consistently derived for all the investigated iCOMs.


\subsection{Gas temperature and column density estimates}\label{subsec:3.2}
\begin{figure*}
    \centering
    \includegraphics[width=\textwidth]{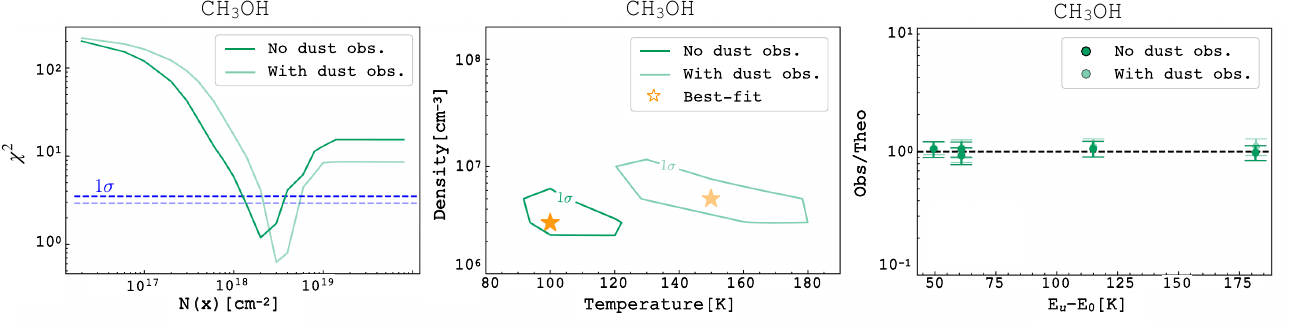}
    \caption{Non--LTE LVG fitting results for methanol (CH$_3$OH) with \textit{(full color)} and without \textit{(shaded)} considering the contribution from mm foreground dust absorption (see Sec. \ref{subsec:3.2}). \textit{Left}: $\chi^2$ minimisation for the total column density N$_{\rm CH_3OH}$(A-type plus E-type).
    \textit{Middle:} Density–temperature $\chi^2$ contour plots showing 1$\sigma$ confidence level, assuming the best-fit value of N$_{\rm CH_3OH}=2\times10^{18}$ cm$^{-2}$ and a source size of $0\farcs3$. The best-fit solutions are marked by the star.
    \textit{Right:} Observed line integrated intensities versus modeled ones as a function of the upper-state energy level value with respect to the lowest value.}
    \label{fig:3}
\end{figure*}

\begin{figure*}[htp!]
    \centering
    \includegraphics[width=\textwidth]{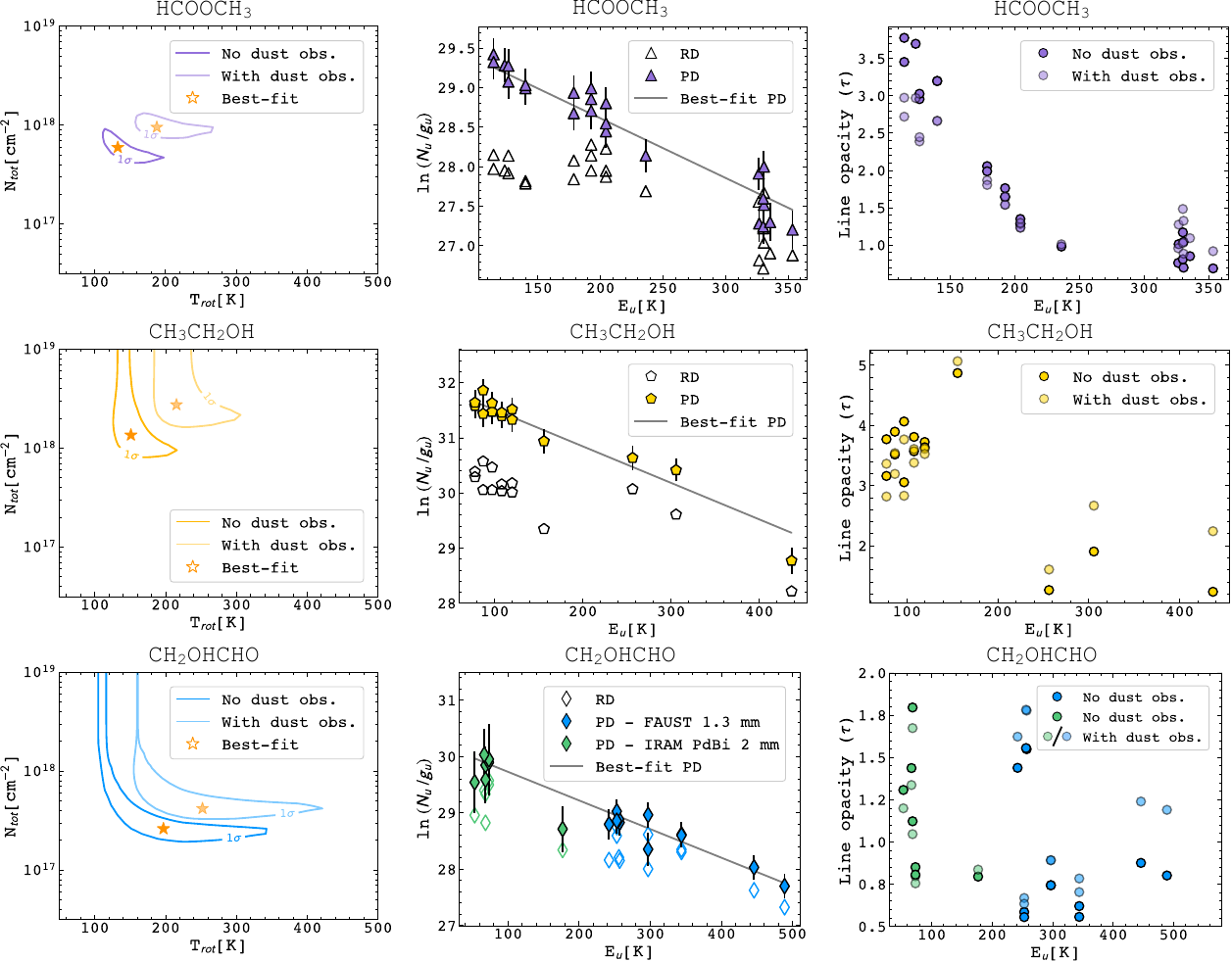}
    \caption{LTE population diagrams (PD) of methyl formate (HCOOCH$_3$), ethanol (CH$_3$CH$_2$OH), and glycolaldehyde (CH$_2$OHCHO). For CH$_2$OHCHO, we complemented our dataset \textit{(blue diamonds)} with the dataset of \citet{taquet_2015} \textit{(green diamonds)}. The results are plotted with \textit{(shaded)} and without \textit{(full color)} the mm dust obscuration factor. \textit{Left:} Column density-temperature reduced $\tilde{\chi}^2$ contour plots. The shaded and solid colored lines indicate the 1$\sigma$ confidence ranges. The star marks the best-fit value in the two cases. \textit{Center:} Observed transitions corrected for the line opacity calculated at the best-fit values. The grey solid line refers to the best-fit of the data points (not corrected for dust obscuration). Empty markers indicate the points not corrected for line opacity, whose fit is the rotational diagram (RD). \textit{Right:} Line opacity as a function of the upper-state energy of the best-fit values.}
    \label{fig:4}
    \vspace{-0.5cm}
\end{figure*}
In order to characterize the hot corino species column density and gas temperature, we analyzed the IRAS 4A2 spectrum performing a Gaussian fit on the emitting iCOMs lines (Fig \ref{fig:6}--\ref{fig:11}). 
Table \ref{tab2} reports the identified lines per species, with their spectral parameters and the results from the Gaussian fits. The line peak velocities lie between $+5.9$ and $+7.5$ km s$^{-1}$, consistent with the systemic velocity of IRAS 4A2 \citep[$+6.8$ km s$^{-1}$;][]{choi_2001} given the channel resolution of $\sim1.4$ km s$^{-1}$. At this spectral resolution we do not detect significant deviations of the line profiles from a thermally broadened Gaussian. 
Note that we excluded from the analysis heavily blended transitions where we could not disentangle the emission. 

The standard Rotational Diagram (RD) method \citep{blake_1987, turner_1991, goldsmith_1999, mangum_2015} assumes local thermodynamic equilibrium (LTE) and optically-thin line emission. While LTE is generally valid due to the high densities of the probed region \citep[higher than the species critical density;][]{marta_2020, marta2022b}, the optical depth assumption is likely not applicable for most of the iCOMs targeted here. This can lead to a potential underestimation of the upper-state populations of these transitions \citep{taquet_2015, marta_2020}.

Methanol (CH$_3$OH) is known to be very abundant and optically-thick in Class 0 sources, including IRAS 4A2. It is also one of the very few complex organic molecules for which collisional coefficients have been computed in order to perform a non-LTE analysis. We therefore performed a non-LTE analysis via our custom large velocity gradient (LVG) code \texttt{grelvg} \citep{ceccarelli_2003, marta_2020}. With this, we fit the observed molecular line intensities and compared them with the predicted values using a chi-square minimization \textbf{($\tilde{\chi}^2$)}, accounting for line opacity. More details on the method are in App. \ref{appendixC}. Table \ref{tab1} and Fig. \ref{fig:3} report the 1$\sigma$ confidence level range for column density and temperature assuming a $0\farcs3$ emitting size, as derived from the CH$_3$OH integrated intensity maps. The resulting reduced chi-square is 0.6. With an LTE rotational diagram, at 1$\sigma$ the resulting column density is constrained to 0.9--1.0 10$^{18}$ cm$^{-2}$, and the temperature to 148--168 K. In comparison, the LTE rotational diagram leads to a lower column density by a factor $\sim$ 2, while overestimating the temperature by a factor $\sim$ 1.3. The detected lines are too few to perform a statistically significant LTE population diagram analysis.

\begin{table*}[htp!]
    \caption{1$\sigma$ confidence level results of the iCOMs molecular lines analysis and line emitting radius fit toward IRAS 4A2.} \label{tab1}
    \centering
    {%
    \begin{tabular}{l|cccc|cc}
    \specialrule{.2em}{.1em}{.1em}
    Species & \multicolumn{4}{c|}{Physical parameters} & \multicolumn{2}{c}{Image fitting}\\
    \hline
    & \multicolumn{2}{c}{No dust obscuration} & \multicolumn{2}{c|}{With dust obscuration} & & \\
    & T$_{\rm gas}$ [K]& N$_{\rm tot}$ [cm$^{-2}$] & T$_{\rm gas}$ [K]& N$_{\rm tot}$ [cm$^{-2}$] & $\bar{r}^{~a}$ [au] & $\theta_{\rm source}^{~b}$ [arcsec]\\
    \noalign{\vskip 0.5mm}
    \hline
    \noalign{\vskip 0.5mm}
    & \multicolumn{4}{c|}{non-LTE LVG} \\
    \hline 
    \noalign{\vskip 0.5mm}
    CH$_3$OH & 90-120 & 1.4-4 $\times$ 10$^{18}$ & 100-140 & 3 $\times$ 10$^{18}$ & 40.1 (0.8) & 0.30\\
    \hline
    \noalign{\vskip 0.5mm}
    & \multicolumn{4}{c|}{LTE population diagram (PD)} \\
    \hline 
    \noalign{\vskip 0.5mm}
    HCOOCH$_3$ & 110-200 & 0.4-1.0 $\times$ 10$^{18}$ & 160-270 & 0.8-1.3 $\times$ 10$^{18}$ & 36.0 (1.0) & 0.26 \\
    CH$_3$CH$_2$OH & 130-215 & $>$8 $\times$ 10$^{17}$ & 190-310 & >1.8 $\times$ 10$^{18}$ & 27.0 (0.8) & 0.19\\
    CH$_2$OHCHO & 110-340 & $>$2 $\times$ 10$^{17}$ & 160-430 & $>$3.5 $\times$ 10$^{17}$ & 22.0 (1.0) & 0.15\\
    \hline
    \noalign{\vskip 0.5mm}
    & \multicolumn{4}{c|}{LTE rotational diagram (RD)} \\
    \hline
    \noalign{\vskip 0.5mm}
    CH$_3$CHO & 180-228 & 2.1-3.1 $\times$ 10$^{17}$ & 180-228 & 3.3-4.7 $\times$ 10$^{17}$ & 28.0 (0.9) & 0.20\\
    NH$_2$CHO & 197$^{~c}$ & 1.4-1.7 $\times$ 10$^{16}$ & 197$^{~c}$ & 2.1-2.5 $\times$ 10$^{16}$ & 23.0 (0.6) & 0.16\\
    \specialrule{.2em}{.1em}{.1em}
    \end{tabular}
    }
    \begin{tablenotes}
    \footnotesize
    \item $^{a}$Weighted average emitting radius. The uncertainties are reported in round parentheses. $^{b}$Source size adopted for the analysis. $^{c}$Fixed as the best-fit temperature derived for CH$_2$OHCHO.
    \end{tablenotes}
    \vspace{-0.5cm}
\end{table*}

For the other iCOMs, lacking collisional coefficients for an LVG analysis, we used where possible the population diagram (PD) approach \citep{goldsmith_1999}, which corrects rotational temperature and column density estimates for the self-consistently estimated line opacity. The methodology is described in detail in App. \ref{appendixD}.
It is important to note that lines with high excitation (E$_{\rm u}\geq$ 300 K) may not be collisionally dominated, but rather radiatively pumped by the infrared field of the protostar \citep{tielens2005}. Therefore, they may not represent the actual kinetic temperature of the gas and may bias the population diagram results. 
Including radiative pumping would require detailed physical and chemical modeling of the envelope that is beyond the scope of this work. 
Because of this, to be on the safe side, at first we restricted the analysis to lines with E$_{\rm u} \leq$ 300 K  (Tab. \ref{tab4} and Fig. \ref{fig:14}), with resulting reduced chi-square at 1$\sigma$ of 0.6 (HCOOCH$_3$), 0.4 (CH$_3$CH$_2$OH), and 0.6 (CH$_2$OHCHO). Then, we repeated the analysis accounting for all detected lines, including the ones with E$_{\rm u}$ $>$ 300 K (Fig. \ref{fig:4}). 
We found that the best-fit results, column densities and temperatures, are consistent between the two cases. Considering all lines, we obtain 1$\sigma$ confidence level ranges with $\tilde{\chi}^2$ values of 1.0 (HCOOCH$_3$), 0.9 (CH$_3$CH$_2$OH), and 0.5 (CH$_2$OHCHO). The best fits to the population diagram (see Fig. \ref{fig:4} and \ref{fig:14}) are similar in the two cases, implying that the bulk of the emission of the two groups of transitions (with E$_{\rm u}$ below and above 300 K) can be represented by the same physical conditions. In light of these results, we conclude that, in our case, the radiative pumping contribution may be negligible and we therefore rely on the results considering all transitions, which also allow more reliable 1$\sigma$ estimates.
Note that for glycolaldehyde (CH$_2$OHCHO) the FAUST data only cover transitions with high E$_{\rm u}$. We therefore complemented our dataset with the sample from \citet{taquet_2015}, as it covers transitions with E$_{\rm u}$ below 200 K, accounting for the different beam size ($\sim2"$). 
We confirm that most of the lines are optically-thick ($\tau>1$), justifying the requirement for the population diagram analysis. 
The acetaldehyde (CH$_3$CHO) transitions are too few to perform a statistically significant PD analysis. Thus, we relied on the RD analysis; the results are reported in Tab. \ref{tab1} and Fig. \ref{fig:15}. 
Finally, the few detected formamide (NH$_2$CHO) lines span a too small range of E$_{\rm u}$ to obtain reliable results from a rotational diagram. Nonetheless, given that NH$_2$CHO traces a similar spatial scale to CH$_2$OHCHO, we derived a lower limit on the column density (see Tab. \ref{tab1}), assuming the same best-fit temperature derived for CH$_2$OHCHO.

\begin{figure}[htp!] 
\centering
\includegraphics[width=0.9\columnwidth]{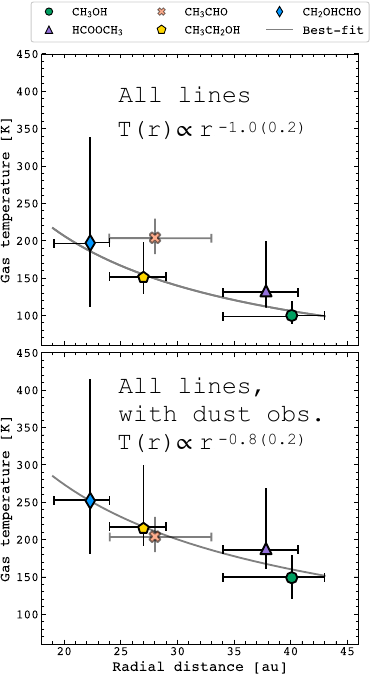}
    \caption{T--r profile of IRAS 4A2. Gas temperatures are derived accounting for all transitions and for the mm dust obscuration \textit{(bottom)}. The colored data points mark the best-fit gas temperature derived for each species as a function of the average emitting radius $\bar{r}$. The emitting radius error bar encompasses the observed range of sizes per species. The error bar on the temperature is the 1$\sigma$ confidence range of the LTE rotational diagram analysis (CH$_3$CHO), LTE population diagram analysis (HCOOCH$_3$, CH$_3$CH$_2$OH, and CH$_2$OHCHO), and of the non-LTE LVG analysis for CH$_3$OH. The grey solid line in both panels is the best fit to the data points excluding CH$_3$CHO (see Sec. \ref{subsec:3.2}).
   }\label{fig:5}
\end{figure}


\section{Discussion} \label{sec:discussion}
\subsection{The effect of millimeter foreground dust absorption}\label{subsec:4.1}
There are several possible mechanisms by which the observed molecular line emission may be obscured or otherwise reduced in the presence of optically-thick dust. 

First, Class 0/I sources may be embedded in envelopes which often are optically-thick at mm/sub-mm wavelengths \citep{miotello_2014, galvan_2018, galametz_2019}. In this case, the foreground dust may absorb the molecular emission, what we call here \textit{millimeter dust obscuration}. This leads to a non-detection of iCOMs at millimeter wavelengths \citep[e.g.][]{lopez_2017, marta_2017, belloche_2020}, or to the underestimation of the molecular column density inferred from optically-thin lines, or an underestimation of both column density and temperature when the optical depth of the lines is not negligible \citep{marta_2020}.
Indeed, even if the dust is co-spatial with the gas, in the case of optically-thick dust and LTE molecules in emission, the line emission will be inseparable from the dust continuum emission. Therefore, the process of continuum subtraction will result in reduced (or apparently absent, or negative) line emission, regardless of the lines being intrinsically optically-thin or optically-thick.

As another possibility, if the molecular gas resides in front of optically-thick dust continuum emission, then after continuum subtraction the molecular lines may appear either in emission or absorption, or not appear at all, depending on the relative brightness temperatures of the line- and the dust-emitting regions.
It can be difficult to distinguish between these scenarios without considering additional information about the source of the emission. Nevertheless, in all cases the derived molecular column densities and excitation temperatures can be significantly underestimated if the dust emission is not taken into account.

\citet{marta_2020, marta_2022} uniquely estimated, using a combination of centimeter and millimeter observations, that the foreground dust may absorb about 30\% at 143 GHz and 50\% at 243 GHz the emission lines in IRAS\,4A2 on scales of $0\farcs26$. Assuming all species on scales $< 0\farcs26$ will be affected in some way by millimeter dust obscuration, we corrected the integrated line flux for these factors (see the results in Tab. \ref{tab1}). In the PD and the LVG analysis, we found a systematic shift toward higher temperatures by 50–100 K and higher column densities by a factor of about 1.6 (2.2 for CH$_3$CH$_2$OH). For acetaldehyde (CH$_3$CHO) and formamide (NH$_2$CHO), we found a similar increase in column density from the RD analysis. 

We compared our estimates to the literature \citep{taquet_2015, marta_2017, marta_2020, lopez_2017, belloche_2020}.
In particular, \citet{belloche_2020} were the first to isolate IRAS 4A2 as a compact hot corino, yet, higher spatial resolution, achieved with this work, was necessary to spatially resolve the emitting size of various iCOMs.
The values derived for methanol (CH$_3$OH) correcting for dust obscuration are consistent at 1$\sigma$ ($\tilde{\chi}^2$ = 0.3) with those derived at centimeter wavelengths by \citet{marta_2020}. The foreground dust absorption affects column density and temperature similarly to the line opacity (respectively with increasing factors of $\sim$ 1.4 and $\sim$ 3), with respect to the LTE rotational diagram method. For the other iCOMs, our new estimates generally point towards higher column densities and higher temperatures than previously measured. These discrepancies are due to  a combination of line optical depth and foreground dust absorption effects.

Taking into account the foreground dust absorption and the measured line opacities, we do not observe systematic differences for linewidths and position angles among the different molecular species (see Tab. \ref{tab2}). Only CH$_3$OH shows slightly larger linewidths, on average $\sim$ 4 km s$^{-1}$ instead of $\sim$ 3.5 km s$^{-1}$ as for the other species. This could be due to the fact that methanol's transitions may trace, particularly at low upper-state energy, other physical processes in the protostellar environment, such as outflowing material \citep[e.g.][]{marta_2024}. From the measured position angles, we estimate an inclination angle in the range of 11–56 degrees. However, the measured linewidths and position angles are uncertain, and the error bars, especially for the latter, are often large. A larger spectral resolution is required to conduct a detailed study of the kinematics of the IRAS 4A system.
\subsection{The gas temperature profile of the IRAS 4A2 hot corino} 
\label{Sec:4.2}

The iCOMs spatial segregation together with the temperature estimate allow us to tentatively derive the gas temperature profile of IRAS 4A2 as a function of radial distance. Fig. \ref{fig:5} shows the derived gas temperatures (with and without accounting for the mm dust obscuration) versus the deconvolved emitting size derived for each species, as a probe of the distance from the protostellar center. For completeness, Fig. \ref{fig:16} shows the resulting temperature profile accounting only for iCOMs transitions with E$_{\rm u}$ $\leq$ 300 K. We naturally excluded from the fit formamide (NH$_2$CHO), for which it was not possible to reliably estimate the rotational temperature, and acetaldehyde (CH$_3$CHO), for which we could not estimate self-consistently the contribution of the line opacity. Nevertheless, we included as a reference the CH$_3$CHO temperature derived with the RD in Fig. \ref{fig:5}.

From theory, the dust temperature is expected to radially behave like a power law: T(r) $\propto$ r$^{~ q}$ \citep{beckwith_1990, motte_2001, andrews_2007}. At large densities, such as in hot corinos, gas and dust are likely thermally coupled \citep{ceccarelli_1996, maret_2002, crimier_2009, crimier_2010}.
By fitting the data points with a power-law function, we find an exponent $q = -1.0 \pm 0.2$. Accounting for the millimeter dust obscuration, we obtain a consistent result of $q= -0.8 \pm 0.2$.
A similar attempt in low mass protostars has been performed at larger scales, focusing on modeling and/or observing simpler molecular species (e.g., CO, H$_2$CO, H$_2$CS) and hinting to a profile of spherically collapsing envelopes \citep{ceccarelli_2000a, maret_2002, crimier_2010}, or rotationally supported disk \citep{jacobsen_2018, vantoff_2019}. 
Our fit implies a power-law steeper than what expected in a optically-thin spherical collapsing envelope \citep[$q\sim$ 0.4--0.5;][]{ceccarelli_2000a, schoier_2002, crimier_2010}. Alternatively, we might be observing layers in the envelope where the dust optical depth is $\leq$ 1 \citep{adams_1986, ceccarelli_1996}. Indeed, from the comparison with observations at centimeter wavelength, \citet{marta_2020} estimated a dust foreground absorption of 30\% circa, which corresponds to a dust optical depth of about 0.3. 
This is reasonable since a completely optically-thick envelope ($\tau$ $>$ 1) would have completely absorbed the molecular line emission, as in the case of the IRAS 4A1 binary component \citep{marta_2020}.
Another possibility is that, to explain the high gas temperature observed at these compact scales, the iCOMs are probing a disk-like structure that is gravitationally unstable \citet{Zamponi_2021}.

To disantangle which contribution dominates, we need a detailed thermochemical model to reproduce the T--r profile, and high angular (about 5--10 au) observations at higher spectral resolution so as to resolve the kinematics of the most compact iCOMs species.

\section{Conclusions} 
\label{sec:conclusions}
Within the FAUST framework we observed for the first time a chemical segregation of six iCOMs in the IRAS 4A2 hot corino.
\begin{enumerate}
    \item The detected iCOMs show different emitting sizes of increasing outer radius, with methanol (CH$_3$OH) being the most extended, at $\sim$ 40 au, glycolaldehyde (CH$_2$OHCHO) and formamide (NH$_2$CHO) the most compact and partially unresolved, at $\sim$ 22--23 au, and methyl formate (HCOOCH$_3$), acetaldehyde (CH$_3$CHO) and ethanol (CH$_3$CH$_2$OH) in between, with an outer radius located at $\sim$ 28--36 au.
    \item Using a multi-line analysis, both in LTE and non-LTE (the latter for CH$_3$OH only), we derived gas temperatures and molecular column densities corrected for (i) line opacity and for (ii) the foreground dust absorption at millimeter wavelengths. The latter implies higher gas temperatures by 50--100 K, and higher column densities by a factor $\sim$ 1.6. 
    \item We retrieved a gas temperature profile (T--r) at scales of 20--50 au (in radius) directly from the observed molecular line emission. The higher gas temperature (up to 200--250 K) is associated with the more compact iCOM emission. The power-law T--r exponent $q$ is about -0.8/-1.0, inconsistent with an optically-thin spherical collapsing envelope. This may hint at a partially optically-thick protostellar envelope, or a gravitationally unstable disk-like structure in IRAS 4A2.
\end{enumerate}
At millimeter wavelengths, the optical thickness of foreground dust and the molecular line opacity are crucial parameters to determine unbiased gas temperatures and molecular abundances. This work highlights how the mm dust opacity affects not only the estimates of the column densities and/or abundances of the observed molecular species, but also the temperature of the traced gas. This is particularly evident for the more abundant species, e.g. methanol (CH$_3$OH), which has optically thick emission lines. However, high spatial and spectral resolution observations, paired with a large bandwidth, are still needed to spatially resolve the molecular emission as well as the gas kinematics, especially of the most compact iCOM species. This will also allow to clarify the exact source geometry.

This work opens the way to many future perspectives. First of all, there is a need to derive similar temperature profiles for other Class 0/I protostars, and compare them with the predictions on embedded young disks from the latest simulations \citep[e.g.][]{Lebreuilly2024}. Also, as mentioned in Sect. \ref{subsec:3.1}, there is a need to link the stratified chemical structure to the iCOMs binding energies, and perform a similar study on other hot corino sources, which also implies follow-ups at higher angular resolution. Such work also sets the ground for the upcoming centimeter facilities (e.g., SKA\footnote{\url{https://www.skao.int/en}} and ngVLA\footnote{\url{https://ngvla.nrao.edu}}), as well as the new ALMA wide band sensitivity upgrade \citep{WSU}. 
Centimeter observations are particularly crucial to fully characterize hot corinos both from a chemical and physical prospective. With SKA and ngVLA, we will access much larger and complex iCOMs species than methanol, in a wavelength regime where the dust is more likely optically-thin, and spectral line blending and line confusion are reduced. At the same time, ALMA WSU will provide us with a large bandwidth to detect more transitions per species and therefore perform a more robust multi-line analysis. 

\begin{acknowledgements} We thank the anonymous referee for all the useful comments and suggestions that greatly improved the manuscript.
This Paper makes use of the following ALMA data: ADS/JAO.ALMA\#2018.1.01205.L (PI: S. Yamamoto). ALMA is a partnership of the ESO (representing its member states), the NSF (USA) and NINS (Japan), together with the NRC (Canada) and the NSC and ASIAA (Taiwan), in cooperation with the Republic of Chile. The Joint ALMA Observatory is operated by the ESO, the AUI/NRAO, and the NAOJ.
This work was partly supported by the Italian Ministero dell Istruzione, Universit\`a e Ricerca through the grant Progetti Premiali 2012 – iALMA (CUP C$52$I$13000140001$).
This project has received funding from the European Union's Horizon 2020 research and innovation programme under the Marie Sklodowska-Curie grant agreement No 823823 (DUSTBUSTERS) and from the European Research Council (ERC) via the ERC Synergy Grant {\em ECOGAL} (grant 855130).
Part of the analysis and work that led to this study were carried out during the May 2023 workshop at the Institut Pascal in Saclay, which was funded and organized as part of the ECOGAL collaboration.
JF acknowledges financial support from the DIFA and the ESO Office for Science. ClCO, LP, GS and EB acknowledge the PRIN-MUR 2020 BEYOND-2p (Astrochemistry beyond the Second period elements, Prot. 2020AFB3FX), the project ASI-Astrobiologia 2023 MIGLIORA (Modeling Chemical Complexity, F83C23000800005), the INAF-GO 2023 fundings PROTO-SKA (Exploiting ALMA data to study planet forming disks: preparing the advent of SKA, C13C23000770005), the INAF Mini-Grant 2022 “Chemical Origins” (PI: L. Podio), the INAF Mini-grant 2023 TRIESTE (“TRacing the chemIcal hEritage of our originS: from proTostars to planEts”; PI: G. Sabatini), and the National Recovery and Resilience Plan (NRRP), Mission 4, Component 2, Investment 1.1, Call for tender No. 104 published on 2.2.2022 by the Italian Ministry of University and Research (MUR), funded by the European Union – NextGenerationEU– Project Title 2022JC2Y93 Chemical Origins: linking the fossil composition of the Solar System with the chemistry of protoplanetary disks – CUP J53D23001600006 - Grant Assignment Decree No. 962 adopted on 30.06.2023 by the Italian Ministry of Ministry of University and Research (MUR). E.B. also acknowledges the contribution of the Next Generation EU funds within the National Recovery and Resilience Plan (PNRR), Mission 4 - Education and Research, Component 2 - From Research to Business (M4C2), Investment Line 3.1 - Strengthening and creation of Research Infrastructures, Project IR0000034 – “STILES - Strengthening the Italian Leadership in ELT and SKA”. I.J-.S acknowledges funding from grant PID2022-136814NB-I00 funded by MICIU/AEI/ 10.13039/501100011033 and by “ERDF/EU”. SBC was supported by the NASA Planetary Science Division Internal Scientist Funding Program through the Fundamental Laboratory Research work package (FLaRe). M.B. acknowledges the support from the European Research Council (ERC) Advanced Grant MOPPEX 833460.
\end{acknowledgements}

\bibliographystyle{aa}
\bibliography{sample631}{}

\begin{thebibliography}{85}
\expandafter\ifx\csname natexlab\endcsname\relax\def\natexlab#1{#1}\fi

\bibitem[{{Adams} \& {Shu}(1986)}]{adams_1986}
{Adams}, F.~C. \& {Shu}, F.~H. 1986, \apj, 308, 836

\bibitem[{{Andr\'e}(2002)}]{andre_2002}
{Andr\'e}, P. 2002, EAS Publications Series, 3, 1

\bibitem[{{Andr\'e} {et~al.}(2000){Andr\'e}, {Ward-Thompson}, \& {Barsony}}]{andre_2000}
{Andr\'e}, P., {Ward-Thompson}, D., \& {Barsony}, M. 2000, in Protostars and Planets IV, ed. V.~{Mannings}, A.~P. {Boss}, \& S.~S. {Russell}, 59

\bibitem[{{Andrews} \& {Williams}(2007)}]{andrews_2007}
{Andrews}, S.~M. \& {Williams}, J.~P. 2007, The Astrophysical Journal, 659, 705

\bibitem[{{Beckwith} {et~al.}(1990){Beckwith}, {Sargent}, {Chini}, \& {Guesten}}]{beckwith_1990}
{Beckwith}, S. V.~W., {Sargent}, A.~I., {Chini}, R.~S., \& {Guesten}, R. 1990, The Astronomical Journal, 99, 924

\bibitem[{{Belloche} {et~al.}(2020){Belloche}, {Maury}, {Maret}, {Anderl}, {Bacmann}, {Andr{\'e}}, {Bontemps}, {Cabrit}, {Codella}, {Gaudel}, {Gueth}, {Lef{\`e}vre}, {Lefloch}, {Podio}, \& {Testi}}]{belloche_2020}
{Belloche}, A., {Maury}, A.~J., {Maret}, S., {et~al.} 2020, \aap, 635, A198

\bibitem[{{Bernab{\`o}} {et~al.}(2022){Bernab{\`o}}, {Turrini}, {Testi}, {Marzari}, \& {Polychroni}}]{Bernabo2022}
{Bernab{\`o}}, L.~M., {Turrini}, D., {Testi}, L., {Marzari}, F., \& {Polychroni}, D. 2022, \apjl, 927, L22

\bibitem[{{Bianchi} {et~al.}(2022){Bianchi}, {L{\'o}pez-Sepulcre}, {Ceccarelli}, {Codella}, {Podio}, {Bouvier}, \& {Enrique-Romero}}]{bianchi_2022}
{Bianchi}, E., {L{\'o}pez-Sepulcre}, A., {Ceccarelli}, C., {et~al.} 2022, \apjl, 928, L3

\bibitem[{{Blake} {et~al.}(1987){Blake}, {Sutton}, {Masson}, \& {Phillips}}]{blake_1987}
{Blake}, G.~A., {Sutton}, E.~C., {Masson}, C.~R., \& {Phillips}, T.~G. 1987, The Astrophysical Journal, 315, 621

\bibitem[{{Bottinelli} {et~al.}(2004){Bottinelli}, Ceccarelli, Lefloch, Williams, Castets, Caux, Cazaux, Maret, Parise, \& Tielens}]{bottinelli_2004a}
{Bottinelli}, S., Ceccarelli, C., Lefloch, B., {et~al.} 2004, The Astrophysical Journal, 615, 354

\bibitem[{{Calcutt} {et~al.}(2014){Calcutt}, {Viti}, {Codella}, {Beltr{\'a}n}, {Fontani}, \& {Woods}}]{calcutt_2014}
{Calcutt}, H., {Viti}, S., {Codella}, C., {et~al.} 2014, \mnras, 443, 3157

\bibitem[{{Caselli} \& {Ceccarelli}(2012)}]{caselli_2012}
{Caselli}, P. \& {Ceccarelli}, C. 2012, A\&ARv, 20, 56

\bibitem[{{Ceccarelli}(2004)}]{ceccarelli_2004}
{Ceccarelli}, C. 2004, in Astronomical Society of the Pacific Conference Series, Vol. 323, Star Formation in the Interstellar Medium: In Honor of David Hollenbach, ed. D.~{Johnstone}, F.~C. {Adams}, D.~N.~C. {Lin}, D.~A. {Neufeeld}, \& E.~C. {Ostriker}, 195

\bibitem[{{Ceccarelli}(2023)}]{ceccarelli_2023}
{Ceccarelli}, C. 2023, in European Conference on Laboratory Astrophysics ECLA2020. The Interplay of Dust, 3--16

\bibitem[{Ceccarelli {et~al.}(2017)Ceccarelli, Caselli, Fontani, Neri, López-Sepulcre, Codella, Feng, Jiménez-Serra, Lefloch, Pineda, Vastel, Alves, Bachiller, Balucani, Bianchi, Bizzocchi, Bottinelli, Caux, Chacón-Tanarro, Choudhury, Coutens, Dulieu, Favre, Hily-Blant, Holdship, Kahane, Al-Edhari, Laas, Ospina, Oya, Podio, Pon, Punanova, Quenard, Rimola, Sakai, Sims, Spezzano, Taquet, Testi, Theulé, Ugliengo, Vasyunin, Viti, Wiesenfeld, \& Yamamoto}]{ceccarelli_2017}
Ceccarelli, C., Caselli, P., Fontani, F., {et~al.} 2017, The Astrophysical Journal, 850, 176

\bibitem[{{Ceccarelli} {et~al.}(2000){Ceccarelli}, Castets, Caux, Hollenbach, Loinard, Molinari, \& Tielens}]{ceccarelli_2000a}
{Ceccarelli}, C., Castets, A., Caux, E., {et~al.} 2000, A\&A, 355, 1129

\bibitem[{{Ceccarelli} {et~al.}(2023){Ceccarelli}, {Codella}, {Balucani}, {Bockelee-Morvan}, {Herbst}, {Vastel}, {Caselli}, {Favre}, {Lefloch}, {Oberg}, \& {Yamamoto}}]{ceccarelli_ppvii}
{Ceccarelli}, C., {Codella}, C., {Balucani}, N., {et~al.} 2023, in Astronomical Society of the Pacific Conference Series, Vol. 534, Protostars and Planets VII, ed. S.~{Inutsuka}, Y.~{Aikawa}, T.~{Muto}, K.~{Tomida}, \& M.~{Tamura}, 379

\bibitem[{{Ceccarelli} {et~al.}(1996){Ceccarelli}, {Hollenbach}, \& {Tielens}}]{ceccarelli_1996}
{Ceccarelli}, C., {Hollenbach}, D.~J., \& {Tielens}, A. G.~G.~M. 1996, The Astrophysical Journal, 471, 400

\bibitem[{{Ceccarelli} {et~al.}(2003){Ceccarelli}, {Maret}, {Tielens}, {Castets}, \& {Caux}}]{ceccarelli_2003}
{Ceccarelli}, C., {Maret}, S., {Tielens}, A.~G.~G.~M., {Castets}, A., \& {Caux}, E. 2003, \aap, 410, 587

\bibitem[{{Chahine} {et~al.}(2024){Chahine}, {Ceccarelli}, {De Simone}, {Chandler}, {Codella}, {Podio}, {L{\'o}pez-Sepulcre}, {Sakai}, {Loinard}, {Bouvier}, {Caselli}, {Vastel}, {Bianchi}, {Cuello}, {Fontani}, {Johnstone}, {Sabatini}, {Hanawa}, {Zhang}, {Aikawa}, {Busquet}, {Caux}, {Dur{\'a}n}, {Herbst}, {M{\'e}nard}, {Segura-Cox}, {Svoboda}, {Balucani}, {Charnley}, {Dulieu}, {Evans}, {Fedele}, {Feng}, {Hama}, {Hirota}, {Isella}, {J{\'\i}menez-Serra}, {Lefloch}, {Maud}, {Maureira}, {Miotello}, {Moellenbrock}, {Nomura}, {Oba}, {Ohashi}, {Okoda}, {Oya}, {Pineda}, {Rimola}, {Sakai}, {Shirley}, {Testi}, {Viti}, {Watanabe}, {Watanabe}, {Zhang}, \& {Yamamoto}}]{chahine_2024}
{Chahine}, L., {Ceccarelli}, C., {De Simone}, M., {et~al.} 2024, \mnras, 531, 2653

\bibitem[{{Choi}(2001)}]{choi_2001}
{Choi}, M. 2001, The Astrophysical Journal, 553, 219

\bibitem[{{Codella} {et~al.}(2021){Codella}, {Ceccarelli}, {Chandler}, {Sakai}, {Yamamoto}, \& {FAUST Team}}]{codella_2021}
{Codella}, C., {Ceccarelli}, C., {Chandler}, C., {et~al.} 2021, Front. astron. space sci, 8, 227

\bibitem[{{Collier} {et~al.}(2021){Collier}, {Krueger}, {Miller}, {Zhao}, {Billinghurst}, \& {Raston}}]{glyco_ref2}
{Collier}, B., {Krueger}, K., {Miller}, I., {et~al.} 2021, \apjs, 253, 40

\bibitem[{{Costain} \& {Dowling}(1960)}]{formamide_ref5}
{Costain}, C.~C. \& {Dowling}, J.~M. 1960, \jcp, 32, 158

\bibitem[{{Crimier} {et~al.}(2010){Crimier}, {Ceccarelli, C.}, {Alonso-Albi, T.}, {Fuente, A.}, {Caselli, P.}, {Johnstone, D.}, {Kahane, C.}, {Lefloch, B.}, {Maret, S.}, {Plume, R.}, {Rizzo, J. R.}, {Tafalla, M.}, {van Dishoeck, E.}, \& {Wyrowski, F.}}]{crimier_2010}
{Crimier}, {Ceccarelli, C.}, {Alonso-Albi, T.}, {et~al.} 2010, A\&A, 516, A102

\bibitem[{Crimier {et~al.}(2009)Crimier, Ceccarelli, Lefloch, \& Faure}]{crimier_2009}
Crimier, N., Ceccarelli, C., Lefloch, B., \& Faure, A. 2009, A\&A, 506, 1229–1241

\bibitem[{{De Simone} {et~al.}(2020{\natexlab{a}}){De Simone}, {Ceccarelli}, {Codella}, {Svoboda}, {Chandler}, {Bouvier}, {Yamamoto}, {Sakai}, {Caselli}, {Favre}, {Loinard}, {Lefloch}, {Liu}, {L{\'o}pez-Sepulcre}, {Pineda}, {Taquet}, \& {Testi}}]{marta_2020}
{De Simone}, M., {Ceccarelli}, C., {Codella}, C., {et~al.} 2020{\natexlab{a}}, \apjl, 896, L3

\bibitem[{{De Simone} {et~al.}(2022{\natexlab{a}}){De Simone}, {Ceccarelli}, {Codella}, {Svoboda}, {Chandler}, {Bouvier}, {Yamamoto}, {Sakai}, {Yang}, {Caselli}, {Lefloch}, {Liu}, {L{\'o}pez-Sepulcre}, {Loinard}, {Pineda}, \& {Testi}}]{marta2022b}
{De Simone}, M., {Ceccarelli}, C., {Codella}, C., {et~al.} 2022{\natexlab{a}}, \apjl, 935, L14

\bibitem[{{De Simone} {et~al.}(2022{\natexlab{b}}){De Simone}, Codella, Ceccarelli, L{\'{o} }pez-Sepulcre, Neri, Rivera-Ortiz, Busquet, Caselli, Bianchi, Fontani, Lefloch, Oya, \& Pineda}]{marta_2022}
{De Simone}, M., Codella, C., Ceccarelli, C., {et~al.} 2022{\natexlab{b}}, Monthly Notices of the Royal Astronomical Society, 512, 5214

\bibitem[{{De Simone} {et~al.}(2020{\natexlab{b}}){De Simone}, {Codella}, {Ceccarelli}, {L{\'o}pez-Sepulcre}, {Witzel}, {Neri}, {Balucani}, {Caselli}, {Favre}, {Fontani}, {Lefloch}, {Ospina-Zamudio}, {Pineda}, \& {Taquet}}]{marta_2020b}
{De Simone}, M., {Codella}, C., {Ceccarelli}, C., {et~al.} 2020{\natexlab{b}}, \aap, 640, A75

\bibitem[{{De Simone} {et~al.}(2017){De Simone}, {Codella}, {Testi}, {Belloche}, {Maury}, {Anderl}, {Andr{\'e}}, {Maret}, \& {Podio}}]{marta_2017}
{De Simone}, M., {Codella}, C., {Testi}, L., {et~al.} 2017, \aap, 599, A121

\bibitem[{{De Simone} {et~al.}(2024){De Simone}, {Podio}, {Chahine}, {Codella}, {Chandler}, {Ceccarelli}, {L{\'o}pez-Sepulcre}, {Loinard}, {Svoboda}, {Sakai}, {Johnstone}, {M{\'e}nard}, {Aikawa}, {Bouvier}, {Sabatini}, {Miotello}, {Vastel}, {Cuello}, {Bianchi}, {Caselli}, {Caux}, {Hanawa}, {Herbst}, {Segura-Cox}, {Zhang}, \& {Yamamoto}}]{marta_2024}
{De Simone}, M., {Podio}, L., {Chahine}, L., {et~al.} 2024, \aap, 686, L13

\bibitem[{Drouin(2017)}]{jpl_doc}
Drouin, B.~J. 2017, Journal of Molecular Spectroscopy, 340, 1

\bibitem[{{Dubernet} {et~al.}(2013){Dubernet}, {Alexander}, {Ba}, {Balakrishnan}, {Balan{\c{c}}a}, {Ceccarelli}, {Cernicharo}, {Daniel}, {Dayou}, {Doronin}, {Dumouchel}, {Faure}, {Feautrier}, {Flower}, {Grosjean}, {Halvick}, {K{\l}os}, {Lique}, {McBane}, {Marinakis}, {Moreau}, {Moszynski}, {Neufeld}, {Roueff}, {Schilke}, {Spielfiedel}, {Stancil}, {Stoecklin}, {Tennyson}, {Yang}, {Vasserot}, \& {Wiesenfeld}}]{dubernet_2013}
{Dubernet}, M.~L., {Alexander}, M.~H., {Ba}, Y.~A., {et~al.} 2013, \aap, 553, A50

\bibitem[{Endres {et~al.}(2016)Endres, Schlemmer, Schilke, Stutzki, \& Müller}]{cdms_doc}
Endres, C.~P., Schlemmer, S., Schilke, P., Stutzki, J., \& Müller, H.~S. 2016, Journal of Molecular Spectroscopy, 327, 95, new Visions of Spectroscopic Databases, Volume II

\bibitem[{{Galametz} {et~al.}(2019){Galametz}, {Maury}, {Valdivia}, {Testi}, {Belloche}, \& {Andr{\'e}}}]{galametz_2019}
{Galametz}, M., {Maury}, A.~J., {Valdivia}, V., {et~al.} 2019, \aap, 632, A5

\bibitem[{{Galv{\'a}n-Madrid} {et~al.}(2018){Galv{\'a}n-Madrid}, {Liu}, {Izquierdo}, {Miotello}, {Zhao}, {Carrasco-Gonz{\'a}lez}, {Lizano}, \& {Rodr{\'\i}guez}}]{galvan_2018}
{Galv{\'a}n-Madrid}, R., {Liu}, H.~B., {Izquierdo}, A.~F., {et~al.} 2018, \apj, 868, 39

\bibitem[{Gieser {et~al.}(2021)Gieser, Beuther, Semenov, Ahmadi, Suri, Möller, Beltrán, Klaassen, Zhang, Urquhart, Henning, Feng, Galván-Madrid, de~Souza~Magalhães, Moscadelli, Longmore, Leurini, Kuiper, Peters, Menten, Csengeri, Fuller, Wyrowski, Lumsden, Sánchez-Monge, Maud, Linz, Palau, Schilke, Pety, Pudritz, Winters, \& Piétu}]{Gieser_2021}
Gieser, C., Beuther, H., Semenov, D., {et~al.} 2021, A\&A, 648, A66

\bibitem[{{Gieser} {et~al.}(2019){Gieser}, {Semenov}, {Beuther}, {Ahmadi}, {Mottram}, {Henning}, {Beltran}, {Maud}, {Bosco}, {Leurini}, {Peters}, {Klaassen}, {Kuiper}, {Feng}, {Urquhart}, {Moscadelli}, {Csengeri}, {Lumsden}, {Winters}, {Suri}, {Zhang}, {Pudritz}, {Palau}, {Menten}, {Galvan-Madrid}, {Wyrowski}, {Schilke}, {S{\'a}nchez-Monge}, {Linz}, {Johnston}, {Jim{\'e}nez-Serra}, {Longmore}, \& {M{\"o}ller}}]{Gieser_2019}
{Gieser}, C., {Semenov}, D., {Beuther}, H., {et~al.} 2019, \aap, 631, A142

\bibitem[{{Goldsmith} \& {Langer}(1999)}]{goldsmith_1999}
{Goldsmith}, P.~F. \& {Langer}, W.~D. 1999, \apj, 517, 209

\bibitem[{{Herbst} \& {van Dishoeck}(2009)}]{vandishoeck_2009}
{Herbst}, E. \& {van Dishoeck}, E.~F. 2009, Annual Reviews of Astronomy and Astrophysics, 47, 427

\bibitem[{Ilyushin {et~al.}(2009)Ilyushin, Kryvda, \& Alekseev}]{methylformate_ref}
Ilyushin, V., Kryvda, A., \& Alekseev, E. 2009, Journal of Molecular Spectroscopy, 255, 32

\bibitem[{{Jacobsen} {et~al.}(2018){Jacobsen}, {J\o{}rgensen, J. K.}, {van der Wiel, M. H. D.}, {Calcutt, H.}, {Bourke, T. L.}, {Brinch, C.}, {Coutens, A.}, {Drozdovskaya, M. N.}, {Kristensen, L. E.}, {M\"uller, H. S. P.}, \& {Wampfler, S. F.}}]{jacobsen_2018}
{Jacobsen}, {J\o{}rgensen, J. K.}, {van der Wiel, M. H. D.}, {et~al.} 2018, A\&A, 612, A72

\bibitem[{Jiménez-Serra {et~al.}(2012)Jiménez-Serra, Zhang, Viti, Martín-Pintado, \& de~Wit}]{Jim_nez_Serra_2012}
Jiménez-Serra, I., Zhang, Q., Viti, S., Martín-Pintado, J., \& de~Wit, W.-J. 2012, The Astrophysical Journal, 753, 34

\bibitem[{{Johansen} {et~al.}(2014){Johansen}, {Blum}, {Tanaka}, {Ormel}, {Bizzarro}, \& {Rickman}}]{Johansen2014}
{Johansen}, A., {Blum}, J., {Tanaka}, H., {et~al.} 2014, in Protostars and Planets VI, ed. H.~{Beuther}, R.~S. {Klessen}, C.~P. {Dullemond}, \& T.~{Henning}, 547--570

\bibitem[{Johnson {et~al.}(2009)Johnson, Lovas, \& Kirchhoff}]{formamide_ref4}
Johnson, D.~R., Lovas, F.~J., \& Kirchhoff, W.~H. 2009, Journal of Physical and Chemical Reference Data, 1, 1011

\bibitem[{{Johnson} {et~al.}(2013){Johnson}, {Sams}, {Profeta}, {Akagi}, {Burling}, {Yokelson}, \& {Williams}}]{glyco_ref1}
{Johnson}, T.~J., {Sams}, R.~L., {Profeta}, L. T.~M., {et~al.} 2013, Journal of Physical Chemistry A, 117, 4096

\bibitem[{{Karska} {et~al.}(2013){Karska}, {Herczeg, G. J.}, {van Dishoeck, E. F.}, {Wampfler, S. F.}, {Kristensen, L. E.}, {Goicoechea, J. R.}, {Visser, R.}, {Nisini, B.}, {San Jos\'e-Garc\'{\i}a, I.}, {Bruderer, S.}, {\'{}Sniady, P.}, {Doty, S.}, {Fedele, D.}, {Yildiz, U. A.}, {Benz, A. O.}, {Bergin, E.}, {Caselli, P.}, {Herpin, F.}, {Hogerheijde, M. R.}, {Johnstone, D.}, {J\o{}rgensen, J. K.}, {Liseau, R.}, {Tafalla, M.}, {van der Tak, F.}, \& {Wyrowski, F.}}]{karska_2013}
{Karska}, {Herczeg, G. J.}, {van Dishoeck, E. F.}, {et~al.} 2013, A\&A, 552, A141

\bibitem[{Kirchhoff \& Johnson(1973)}]{formamide_ref3}
Kirchhoff, W.~H. \& Johnson, D.~R. 1973, Journal of Molecular Spectroscopy, 45, 159

\bibitem[{Kleiner {et~al.}(1996)Kleiner, Lovas, \& Godefroid}]{acetaldehyde_ref}
Kleiner, I., Lovas, F.~J., \& Godefroid, M. 1996, Journal of Physical and Chemical Reference Data, 25, 1113

\bibitem[{{Kristensen} {et~al.}(2012){Kristensen}, {van Dishoeck, E. F.}, {Bergin, E. A.}, {Visser, R.}, {Yildiz, U. A.}, {San Jose-Garcia, I.}, {J\o{}rgensen, J. K.}, {Herczeg, G. J.}, {Johnstone, D.}, {Wampfler, S. F.}, {Benz, A. O.}, {Bruderer, S.}, {Cabrit, S.}, {Caselli, P.}, {Doty, S. D.}, {Harsono, D.}, {Herpin, F.}, {Hogerheijde, M. R.}, {Karska, A.}, {van Kempen, T. A.}, {Liseau, R.}, {Nisini, B.}, {Tafalla, M.}, {van der Tak, F.}, \& {Wyrowski, F.}}]{kristensen_2012}
{Kristensen}, {van Dishoeck, E. F.}, {Bergin, E. A.}, {et~al.} 2012, A\&A, 542, A8

\bibitem[{Kukolich \& Nelson(1971)}]{formamide_ref2}
Kukolich, S. \& Nelson, A. 1971, Chemical Physics Letters, 11, 383

\bibitem[{Kurland \& Wilson(2004)}]{formamide_ref1}
Kurland, R.~J. \& Wilson, E.~Bright, J. 2004, The Journal of Chemical Physics, 27, 585

\bibitem[{{Lada}(1987)}]{lada_1987}
{Lada}, C.~J. 1987, in Star Forming Regions, ed. M.~{Peimbert} \& J.~{Jugaku}, Vol. 115, 1

\bibitem[{{Lebreuilly} {et~al.}(2021){Lebreuilly}, {Hennebelle}, {Colman}, {Commer{\c{c}}on}, {Klessen}, {Maury}, {Molinari}, \& {Testi}}]{lebreuilly_protoplanetary_2021}
{Lebreuilly}, U., {Hennebelle}, P., {Colman}, T., {et~al.} 2021, \apjl, 917, L10

\bibitem[{{Lebreuilly} {et~al.}(2023){Lebreuilly}, {Hennebelle}, {Colman}, {Maury}, {Tung}, {Testi}, {Klessen}, {Molinari}, {Commer{\c{c}}on}, {Gonz{\'a}lez}, {Pacetti}, {Somigliana}, \& {Rosotti}}]{Lebreuilly2024}
{Lebreuilly}, U., {Hennebelle}, P., {Colman}, T., {et~al.} 2023, arXiv e-prints, arXiv:2310.19672

\bibitem[{{Lebreuilly} {et~al.}(2024){Lebreuilly}, {Hennebelle}, {Colman}, {Maury}, {Tung}, {Testi}, {Klessen}, {Molinari}, {Commer{\c{c}}on}, {Gonz{\'a}lez}, {Pacetti}, {Somigliana}, \& {Rosotti}}]{lebreuilly_synthetic_2024}
{Lebreuilly}, U., {Hennebelle}, P., {Colman}, T., {et~al.} 2024, \aap, 682, A30

\bibitem[{{Lee} {et~al.}(2022){Lee}, {Codella}, {Ceccarelli}, \& {L{\'o}pez-Sepulcre}}]{lee_2022}
{Lee}, C.-F., {Codella}, C., {Ceccarelli}, C., \& {L{\'o}pez-Sepulcre}, A. 2022, \apj, 937, 10

\bibitem[{{L\'opez-Sepulcre} {et~al.}(2017){L\'opez-Sepulcre}, {Sakai, N.}, {Neri, R.}, {Imai, M.}, {Oya, Y.}, {Ceccarelli, C.}, {Higuchi, A. E.}, {Aikawa, Y.}, {Bottinelli, S.}, {Caux, E.}, {Hirota, T.}, {Kahane, C.}, {Lefloch, B.}, {Vastel, C.}, {Watanabe, Y.}, \& {Yamamoto, S.}}]{lopez_2017}
{L\'opez-Sepulcre}, {Sakai, N.}, {Neri, R.}, {et~al.} 2017, A\&A, 606, A121

\bibitem[{{Manara} {et~al.}(2018){Manara}, {Morbidelli}, \& {Guillot}}]{Manara2018}
{Manara}, C.~F., {Morbidelli}, A., \& {Guillot}, T. 2018, \aap, 618, L3

\bibitem[{Mangum \& Shirley(2015)}]{mangum_2015}
Mangum, J.~G. \& Shirley, Y.~L. 2015, Publications of the Astronomical Society of the Pacific, 127, 266

\bibitem[{{Maret} {et~al.}(2002){Maret}, {Ceccarelli}, {Caux}, {Tielens}, \& {Castets}}]{maret_2002}
{Maret}, S., {Ceccarelli}, C., {Caux}, E., {Tielens}, A.~G.~G.~M., \& {Castets}, A. 2002, A\&A, 395, 573

\bibitem[{{Maureira} {et~al.}(2022){Maureira}, {Gong}, {Pineda}, {Liu}, {Silsbee}, {Caselli}, {Zamponi}, {Segura-Cox}, \& {Schmiedeke}}]{maureira_2022}
{Maureira}, M.~J., {Gong}, M., {Pineda}, J.~E., {et~al.} 2022, \apjl, 941, L23

\bibitem[{{Maury} {et~al.}(2014){Maury}, {Belloche}, {Andr{\'e}}, {Maret}, {Gueth}, {Codella}, {Cabrit}, {Testi}, \& {Bontemps}}]{maury_2014}
{Maury}, A.~J., {Belloche}, A., {Andr{\'e}}, P., {et~al.} 2014, \aap, 563, L2

\bibitem[{{Miotello} {et~al.}(2014){Miotello}, {Testi}, {Lodato}, {Ricci}, {Rosotti}, {Brooks}, {Maury}, \& {Natta}}]{miotello_2014}
{Miotello}, A., {Testi}, L., {Lodato}, G., {et~al.} 2014, \aap, 567, A32

\bibitem[{{Motte} \& {Andr\'e}(2001)}]{motte_2001}
{Motte} \& {Andr\'e}. 2001, A\&A, 365, 440

\bibitem[{{M{\"u}ller} {et~al.}(2005){M{\"u}ller}, {Schl{\"o}der}, {Stutzki}, \& {Winnewisser}}]{muller_2005}
{M{\"u}ller}, H. S.~P., {Schl{\"o}der}, F., {Stutzki}, J., \& {Winnewisser}, G. 2005, Journal of Molecular Structure, 742, 215

\bibitem[{{{\"O}berg} \& {Bergin}(2021)}]{oberg_2021}
{{\"O}berg}, K.~I. \& {Bergin}, E.~A. 2021, Phys. Rep., 893, 1

\bibitem[{{Okoda} {et~al.}(2022){Okoda}, {Oya}, {Imai}, {Sakai}, {Watanabe}, {L{\'o}pez-Sepulcre}, {Saigo}, \& {Yamamoto}}]{okoda_2022}
{Okoda}, Y., {Oya}, Y., {Imai}, M., {et~al.} 2022, \apj, 935, 136

\bibitem[{{Ortiz-Le{\'o}n} {et~al.}(2018){Ortiz-Le{\'o}n}, {Loinard}, {Dzib}, {Kounkel}, {Galli}, {Tobin}, {Evans}, {Hartmann}, {Rodr{\'\i}guez}, {Brice{\~n}o}, {Torres}, \& {Mioduszewski}}]{ortiz_2018}
{Ortiz-Le{\'o}n}, G.~N., {Loinard}, L., {Dzib}, S.~A., {et~al.} 2018, \apjl, 869, L33

\bibitem[{Ossenkopf-Okada {et~al.}(2023)Ossenkopf-Okada, Schaaf, Breloy, \& Stutzki}]{WSU}
Ossenkopf-Okada, V., Schaaf, R., Breloy, I., \& Stutzki, J. 2023, Physics and chemistry of star formation : the dynamical ISM across time and spatial scales : proceedings of the 7th Chile-Cologne-Bonn-Symposium

\bibitem[{Pearson {et~al.}(2008)Pearson, Brauer, \& Drouin}]{ethanol_ref}
Pearson, J.~C., Brauer, C.~S., \& Drouin, B.~J. 2008, Journal of Molecular Spectroscopy, 251, 394, special issue dedicated to the pioneering work of Drs. Edward A. Cohen and Herbert M. Pickett on spectroscopy relevant to the Earth’s atmosphere and astrophysics

\bibitem[{Pickett {et~al.}(1998)Pickett, POYNTER, COHEN, DELITSKY, PEARSON, \& MÜLLER}]{pickett_1998}
Pickett, H., POYNTER, R., COHEN, E., {et~al.} 1998, Journal of Quantitative Spectroscopy and Radiative Transfer, 60, 883

\bibitem[{{Rabli} \& {Flower}(2010)}]{rabli_rotational_2010}
{Rabli}, D. \& {Flower}, D.~R. 2010, \mnras, 406, 95

\bibitem[{{Sch{\"o}ier} {et~al.}(2002){Sch{\"o}ier}, {J{\o}rgensen}, {van Dishoeck}, \& {Blake}}]{schoier_2002}
{Sch{\"o}ier}, F.~L., {J{\o}rgensen}, J.~K., {van Dishoeck}, E.~F., \& {Blake}, G.~A. 2002, A\&A, 390, 1001

\bibitem[{{Sheehan} \& {Eisner}(2018)}]{sheehan_2018}
{Sheehan}, P.~D. \& {Eisner}, J.~A. 2018, ApJ, 857, 18

\bibitem[{Taquet {et~al.}(2015)Taquet, López-Sepulcre, Ceccarelli, Neri, Kahane, \& Charnley}]{taquet_2015}
Taquet, V., López-Sepulcre, A., Ceccarelli, C., {et~al.} 2015, The Astrophysical Journal, 804, 81

\bibitem[{{Tielens}(2005)}]{tielens2005}
{Tielens}, A.~G.~G.~M. 2005, {The Physics and Chemistry of the Interstellar Medium}

\bibitem[{Tung {et~al.}(2024)Tung, Testi, Lebreuilly, Hennebelle, Maury, Klessen, Cacciapuoti, González, Rosotti, \& Molinari}]{tung_2024}
Tung, N.-D., Testi, L., Lebreuilly, U., {et~al.} 2024, The accuracy of ALMA estimates of young disk radii and masses. Predicted observations from numerical simulations

\bibitem[{{Turner}(1991)}]{turner_1991}
{Turner}, B.~E. 1991, The Astrophysical Journal, Supplement, 76, 617

\bibitem[{{Tychoniec} {et~al.}(2020){Tychoniec}, {Manara, Carlo F.}, {Rosotti, Giovanni P.}, {van Dishoeck, Ewine F.}, {Cridland, Alexander J.}, {Hsieh, Tien-Hao}, {Murillo, Nadia M.}, {Segura-Cox, Dominique}, {van Terwisga, Sierk E.}, \& {Tobin, John J.}}]{tychoniec_2020}
{Tychoniec}, L., {Manara, Carlo F.}, {Rosotti, Giovanni P.}, {et~al.} 2020, A\&A, 640, A19

\bibitem[{{van 't Hoff} {et~al.}(2020){van 't Hoff}, {van Dishoeck, Ewine F.}, {J\o{}rgensen, Jes K.}, \& {Calcutt, Hannah}}]{vantoff_2019}
{van 't Hoff}, M. L.~R., {van Dishoeck, Ewine F.}, {J\o{}rgensen, Jes K.}, \& {Calcutt, Hannah}. 2020, A\&A, 633, A7

\bibitem[{Xu {et~al.}(2008)Xu, Fisher, Lees, Shi, Hougen, Pearson, Drouin, Blake, \& Braakman}]{methanol_ref}
Xu, L.-H., Fisher, J., Lees, R., {et~al.} 2008, Journal of Molecular Spectroscopy, 251

\bibitem[{Zamponi {et~al.}(2021)Zamponi, Maureira, Zhao, Liu, Ilee, Forgan, \& Caselli}]{Zamponi_2021}
Zamponi, J., Maureira, M.~J., Zhao, B., {et~al.} 2021, Monthly Notices of the Royal Astronomical Society, 508, 2583–2599

\bibitem[{{Zucker} {et~al.}(2018){Zucker}, {Schlafly}, {Speagle}, {Green}, {Portillo}, {Finkbeiner}, \& {Goodman}}]{zucker_2018}
{Zucker}, C., {Schlafly}, E.~F., {Speagle}, J.~S., {et~al.} 2018, \apj, 869, 83

\end{thebibliography}

\begin{appendix}
\onecolumn
\section{Detected lines, spectral parameters and fit results}\label{appendixA}

\begin{figure*}[htp!]
    \centering
    \includegraphics[width=0.9\textwidth]{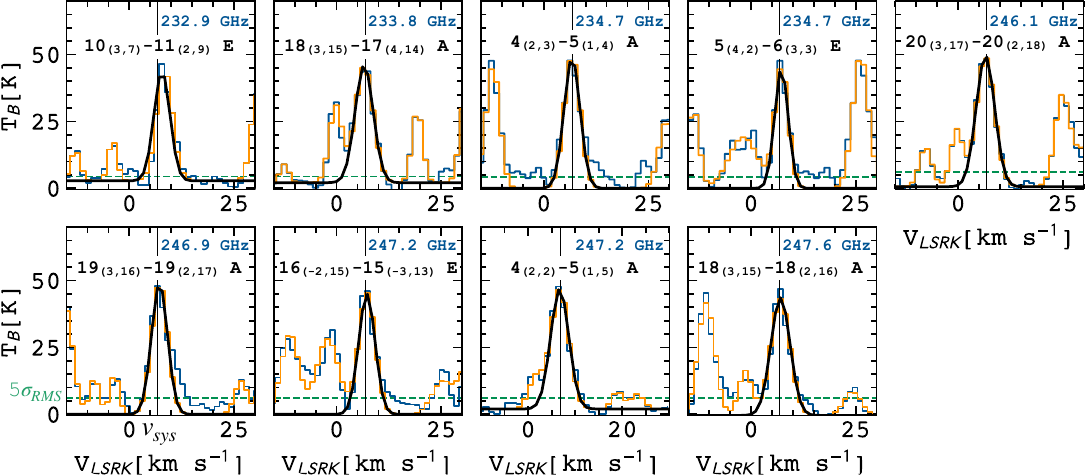}
    \caption{Observed CH$_3$OH line spectra (in T$_B$ scale) extracted at the continuum peak position of IRAS 4A2. The blue solid line indicates the baseline-subtracted spectrum, the orange line the best-fit model, and the black line the Gaussian profile of the labeled transition. The green dotted line marks the 5$\sigma$ line cube RMS. The vertical black solid line indicates the system cloud LSR velocity (+6.8 km s$^{-1}$). The limits on the x-axis are occasionally redefined for visualization purposes.}
\label{fig:6}

\end{figure*}

\begin{figure*}[htp!]
    \centering
    \includegraphics[width=0.56\textwidth]{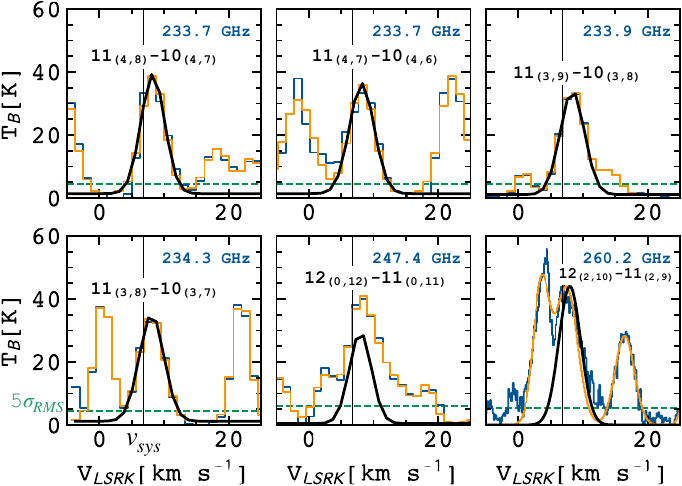}
    \caption{Same as Fig. \ref{fig:6} for NH$_2$CHO.}
\label{fig:7}

\end{figure*}

\begin{figure*}[htp!]
\centering
    \includegraphics[width=\textwidth]{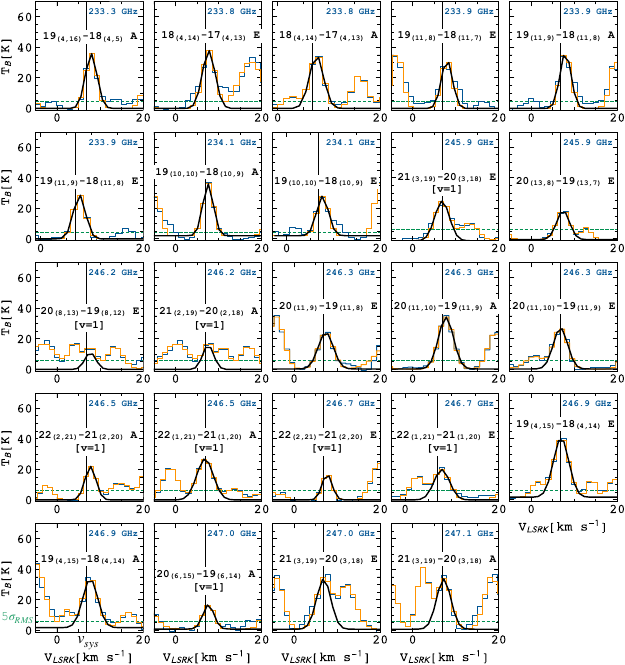}
    \caption{Same as Fig. \ref{fig:6} for HCOOCH$_3$. The limits on the x-axis are occasionally redefined for visualization purposes.}
\label{fig:8}
\end{figure*}

\begin{figure*}[htp!]
    \centering
    \includegraphics[width=\textwidth]{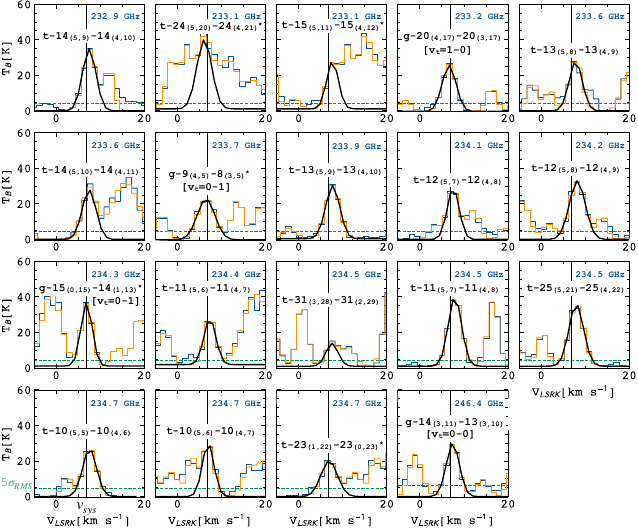}
    \caption{Same as Fig. \ref{fig:6} for CH$_3$CH$_2$OH. Transitions with an asterisk have been excluded from the analysis due to contamination from other species.}
\label{fig:9}
\end{figure*}

\begin{figure*}[htp!]
    \centering
    \includegraphics[width=\textwidth]{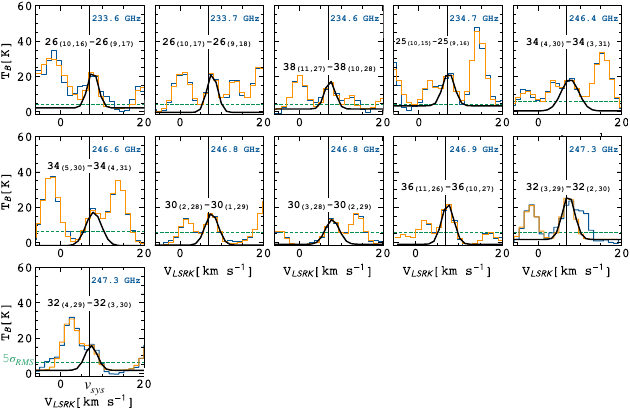}
    \caption{Same as Fig. \ref{fig:6} for CH$_2$OHCHO. The limits on the x-axis are occasionally redefined for visualization purposes.}
\label{fig:10}
\end{figure*}

\begin{figure*}[htp!]
    \centering
    \includegraphics[width=\textwidth]{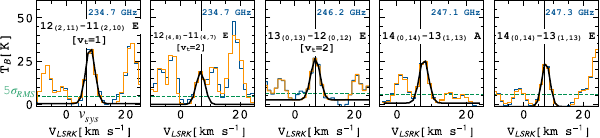}
    \caption{Same as Fig. \ref{fig:6} for CH$_3$CHO. The limits on the x-axis are occasionally redefined for visualization purposes.}
\label{fig:11}
\end{figure*}

\onecolumn
\begin{longtable}{lccccccc}
\caption{Spectral properties and Gaussian fit parameters of iCOMs lines detected towards the IRAS 4A2 continuum peak extracted spectrum.}\label{tab2}\\  
\specialrule{.2em}{.1em}{.1em}
Transition$^{a}$ & \multicolumn{4}{c}{Spectroscopic parameters} & \multicolumn{3}{c}{Gaussian fit results}\\
\cmidrule(r){2-5}
\cmidrule(r){6-8}
& $\nu^{a}$ & E$_{\rm u}^{a}$ & Log$_{10}$ (A$_{\rm ul}$/s$^{-1}$)$^{a}$ & g$_{\rm u}^{a}$ & $\int$T$_{B}$~dV$^{b}$ & V$_{\rm peak}^{b}$ & FWHM$^{b}$ \\
& [MHz] & [K] & & & $\rm [K \ km~ s^{-1}]$ &  [km s$^{-1}$] &  [km s$^{-1}$] \\
\noalign{\vskip 1mm}
\hline
\endfirsthead
\caption{continued.}\\
\endhead
\multicolumn{8}{c}{CH$_3$OH} \\

\hline
10$_{(3,7)}-11_{(2,9)}$ E & 232945.797 & 191 & -4.7 & 84 & 196 (10) & 7.19 (0.10) & 4.5 (0.3) \\
18$_{(3,15)}-17_{(4,14)}$ A & 233795.666 & 447 & -4.7 & 148 & 237 (11) & 5.99 (0.10) & 5.1 (0.3)\\
4$_{(2,3)}-5_{(1,4)}$ A & 234683.370 & 61 & -4.7 & 36 & 237 (11) & 6.16 (0.13) & 4.5 (-)\\
5$_{(4,2)}-6_{(3,3)}$ E & 234698.519 & 123 & -5.2 & 44 & 183 (15) & 6.67 (0.14) & 3.8 (0.3)\\
20$_{(3,17)}$-20$_{(2,18)}$ A & 246074.605 & 537 & -4.1 & 164 & 267 (11) & 5.91 (0.08)  & 5.2 (0.2)\\
19$_{(3,16)}-19_{(2,17)}$ A & 246873.301 & 491 & -4.1 & 156 & 236 (11) & 6.61 (0.13) & 4.5 (-) \\
16$_{(-2,15)}-15_{(-3,13)}$ E & 247161.950 & 338 & -4.6 & 132 & 224 (19) & 6.65 (0.19) & 4.6 (0.5) \\
4$_{(2,2)}-5_{(1,5)}$ A & 247228.587 & 61 & -4.7 & 36 & 217 (8) & 6.12 (0.07) & 4.55 (0.19) \\
18$_{(3,15)}-18_{(2,16)}$ A & 247610.918 & 447 & -4.1 & 148 & 250 (15) & 6.46 (0.13) & 5.4 (0.3)\\

\noalign{\vskip 0.5mm}
\hline 
\multicolumn{8}{c}{HCOOCH$_3$} \\

\hline 
19$_{(4,16)}$-18$_{(4,15)}$ A & 233226.788 & 123 & -3.7 & 78 & 121 (10) & 7.34 (0.13) & 3.2 (0.3)\\
18$_{(4,14)}$-17$_{(4,13)}$ E & 233753.960 & 114 & -3.8 & 74 & 141 (8) & 7.10 (0.12) & 3.5 (-)\\
18$_{(4,14)}$-17$_{(4,13)}$ A & 233777.521 & 114 & -3.8 & 74 & 118 (10) & 6.97 (0.13) & 3.2 (0.3)\\
19$_{(11,8)}$-18$_{(11,7)}$ E & 233845.233 & 193 & -3.9 & 78 & 103 (9) & 7.36 (0.15) & 3.0 (-)\\
19$_{(11,9)}$-18$_{(11,8)}$ A & 233854.286 & 193 & -3.9 & 78 & 118 (9) & 7.36 (0.13) & 3.0 (-)\\
19$_{(11,9)}$-18$_{(11,8)}$ E & 233867.193 & 193 & -3.9 & 78 & 85 (10) & 6.98 (0.16) & 2.8 (0.4)\\ 
19$_{(10,10)}$-18$_{(10,9)}$ A & 234124.883 & 179 & 3.9 & 78 & 105 (9) & 6.95 (0.11) & 2.9 (0.3)\\
19$_{(10,10)}$-18$_{(10,9)}$ E & 234134.600 & 179 & -3.9 & 78 & 83 (8) & 7.07 (0.15) & 3.0 (-)\\
21$_{(3,19)}$-20$_{(3,18)}$ E [v=1] & 245846.914 & 327 & -3.7 & 86 & 57 (18) & 6.24 (0.16) & 4.0 (-)\\
20$_{(13,8)}$-19$_{(13,7)}$ E & 245903.680 & 236 & -3.9 & 82 & 71 (4) & 6.80 (0.07) & 3.5 (0.19)\\
20$_{(8,13)}$-19$_{(8,12)}$ E [v=1] & 246184.177 & 354 & -3.7 & 82 &  46 (6) & 7.14 (-) & 3.0 (-)\\
21$_{(2,19)}$-20$_{(2,18)}$ A [v=1] & 246187.016 & 327 & -3.7 & 86 & 52 (6) & 6.97 (-) & 3.0 (-)\\
20$_{(11,9)}$-19$_{(11,8)}$ E & 246285.400 &  204 & -3.8 & 82 & 103 (8) & 6.75 (0.13) & 4.1 (0.3)\\
20$_{(11,10)}$-19$_{(11,9)}$ A & 246295.135 &  204 & -3.8 & 82 & 147 (8) & 7.13 (0.08) & 4.1 (0.2)\\
20$_{(11,10)}$-19$_{(11,9)}$ E & 246308.272 & 204 & -3.8 & 82 & 112 (8) & 6.20 (0.12) & 3.9 (0.3)\\
22$_{(2,21)}$-21$_{(2,20)}$ A [v=1] & 246461.167 &  331 & -3.7 & 90 & 69 (4) & 7.17 (0.11) & 3.0 (0.3)\\
22$_{(1,21)}$-21$_{(1,20)}$ A [v=1] & 246488.433 &  331 & -3.7 & 90 & 130 (6) & 6.24 (0.09) & 4.5 (0.2)\\  
22$_{(2,21)}$-21$_{(2,20)}$ E [v=1] & 246706.504 &  330 & -3.7 & 90 & 50 (5) & 7.06 (0.13) & 2.7 (0.3)\\
22$_{(1,21)}$-21$_{(1,20)}$ E [v=1] & 246731.729 &  330 & -3.7 & 90 & 83 (9) & 7.23 (0.19) & 3.9 (0.5)\\
19$_{(4,15)}$-18$_{(4,14)}$ E & 246891.611 & 126 & -3.7 & 78 & 176 (11) & 6.35 (0.10) & 4.2 (0.3)\\
19$_{(4,15)}$-18$_{(4,14)}$ A & 246914.658 &  126 & -3.7 & 78 & 141 (11) & 7.12 (0.12) & 4.1 (0.3)\\
20$_{(6,15)}$-19$_{(6,14)}$ A [v=1] & 246985.225 & 336 & -3.7 & 82 & 51 (7) & 7.04 (0.17) & 3.0 (0.4)\\
21$_{(3,19)}$-20$_{(3,18)}$ E & 247044.146 & 140 & -3.7 & 86 & 138 (9) & 6.27 (0.16) & 3.9 (-)\\
21$_{(3,19)}$-20$_{(3,18)}$ A & 247053.453 & 140 & -3.7 & 86 & 138 (8) & 6.61 (0.14) & 3.9 (-)\\
        \noalign{\vskip 0.5mm}
        \hline
        \multicolumn{8}{c}{CH$_3$CHO} \\
        \hline
        12$_{(2,11)}$-11$_{(2,10)}$ E [$v_t$=1] & 233048.516 & 285 & -3.4 & 50 & 126 (9) & 7.52 (0.11) & 3.6 (0.3)\\
        12$_{(4,8)}$-11$_{(4,7)}$ E [$v_t$=2] & 234707.132 & 487 & -3.4 & 50 & 71 (10) & 6.1 (0.3) & 3.5 (-)\\

        13$_{(0,13)}$-12$_{(0,12)}$ E [$v_t$=2] & 246169.217 & 461 & -3.3 & 54 & 95 (8) & 6.29 (0.12) & 3.7 (0.3)\\
        14$_{(0,14)}$-13$_{(1,13)}$ A & 247142.155 & 96 & -4.2 & 58 & 85 (4) & 7.16 (0.05) & 3.1 (0.1)\\
        14$_{(0,14)}$-13$_{(1,13)}$ E & 247341.332 & 96 & -4.2 & 58 & 89 (3) & 6.84 (0.05) & 3.2 (0.1)\\
        \noalign{\vskip 0.5mm}
        \hline
        \multicolumn{8}{c}{NH$_2$CHO} \\
        \hline
        11$_{(4,8)}$-10$_{(4,7)}$ & 233735.603 & 115 & -3.1 & 69 & 174 (13) & 7.66 (0.13) & 4.3 (0.3)\\
        11$_{(4,7)}$-10$_{(4,6)}$ & 233746.504 & 115 & -3.1 & 69 & 165 (10) & 7.58 (0.14) & 4.5 (-)\\
        11$_{(3,9)}$-10$_{(3,8)}$ & 233897.318 & 94 & -3.1 & 69 & 157 (6) & 7.76 (0.08) & 4.41 (0.19)\\
        11$_{(3,8)}$-10$_{(3,7)}$ & 234316.254 & 94 & -3.1 & 69 & 160 (14) & 7.48 (0.14) & 4.5 (0.4)\\
        12$_{(0,12)}$-11$_{(0,11)}$ & 247391.356 & 78 & -3.0 & 75 & 133 (109) & 7.37 (0.18) & 4 (1)\\
        12$_{(2,10)}$-11$_{(2,9)}^{c}$ & 260189.848 & 92 & -2.9 & 75 & 186 (5) & 7.37 (-) & 4 (-)\\
        \noalign{\vskip 0.5mm}
        \hline
        \multicolumn{8}{c}{CH$_3$CH$_2$OH} \\
        \hline
        t-14$_{(5,9)}$-14$_{(4,10)}$ & 232928.499 & 120 & -4.1 & 29 & 133 (10) & 6.91 (0.14) & 3.6 (0.3)\\
        t-24$_{(5,20)}$-24$_{(4,21)}$ & 233208.521 &  285 & -4.1 & 49 & 168 (12) & 5.50 (0.17) & 4.0 (-)\\
        t-15$_{(5,11)}$-15$_{(4,12)}$ & 233215.475 & 132 & -4.1 & 31 & 88 (9) & 7.08 (0.14) & 3.1 (0.3)\\
        g-20$_{(4,17)}$-20$_{(3,17)}$ [$v_t$=1-0] & 233095.873 & 256 & -4.4 & 41 & 103 (10) & 6.03 (0.15) & 3.7 (0.4)\\
        t-13$_{(5,8)}$-13$_{(4,9)}$ & 233571.024 & 108 & -4.2 & 27 & 106 (7) & 7.24 (0.14) & 3.5 (-)\\
        t-14$_{(5,10)}$-14$_{(4,11)}$ & 233601.523 & 120 & -4.1 & 29 & 113 (12) & 6.89 (0.17) & 3.7 (0.4)\\
        g-9$_{(4,5)}$-8$_{(3,5)}$ [$v_t$=0-1] & 233714.281 & 114 & -4.4 & 19 & 117 (9) & 5.90 (0.14) & 4.7 (0.4)\\
        t-13$_{(5,9)}$-13$_{(4,10)}$ & 233951.119 & 108 & -4.1 & 27 & 120 (6) & 7.13 (0.08) & 3.74 (0.19)\\
        t-12$_{(5,7)}$-12${(4,8)}$ & 234051.119 & 97 &  -4.2 & 25 & 98 (8) & 6.95 (0.19) & 3.2 (0.3)\\
        t-12$_{(5,8)}$-12$_{(4,9)}$ & 234255.161 &  97 & -4.2 & 25 & 148 (5) & 7.61 (0.07) & 4.26 (0.17)\\
        g-15$_{(0,15)}$-14$_{(1,13)}$ [$v_t$=0-1] & 234360.555 &  158 & -4.4 & 31 & 115 (9) & 6.31 (0.09) & 3.0 (0.2)\\ 
       t-11$_{(5,6)}$-11$_{(4,7)}$ & 234406.451 & 87 & -4.2 & 23 & 88 (11) & 6.77 (0.17) & 3.2 (0.4)\\
       t-31$_{(3,28)}$-31$_{(2,29)}$ & 234481.343 & 437 & -4.1 &  63 & 45 (6) & 7.0 (-) & 3.2 (0.5)\\
       t-11$_{(5,7)}$-11$_{(4,8)}$ & 234509.565 &  87 &  -4.2 & 23 & 148 (6) & 7.34 (0.07) & 3.55 (0.16)\\
       t-25$_{(5,21)}$-25$_{(4,22)}$ & 234523.893 &  306 & -4.1 & 51 & 143 (6) & 7.37 (0.08) & 3.89 (0.19)\\
       t-10$_{(5,5)}$-10$_{(4,6)}$ & 234666.273 &  78 &  -4.2 & 21 & 108 (10) & 6.99 (0.17) & 3.7 (0.4)\\
       t-10$_{(5,6)}$-10$_{(4,7)}$ & 234714.571 & 78 &  -4.2 & 21 & 98 (10) & 6.4 (-) & 3.1 (0.4)\\ 
       t-23$_{(1,22)}$-23$_{(0,23)}$ & 234725.600 & 233 & -3.8 & 47 & 93 (11) & 6.3 (0.3) & 4.4 (0.6)\\
       g-14$_{(3,11)}$-13$_{(3,10)}$ [$v_t$=0-0] &  246414.762 & 156 & -3.9 & 29 & 110 (10) & 6.88 (0.15) & 3.5 (0.4)\\
        \noalign{\vskip 0.5mm}
        \hline
        \multicolumn{8}{c}{CH$_2$OHCHO} \\
        \hline
        26$_{(10,16)}$-26$_{(9,17)}$ & 233587.449 & 256 & -3.7 & 53 & 68 (9) & 7.1 (0.2) & 3.0 (-)\\
        26$_{(10,17)}$-26$_{(9,18)}$ & 233709.657 & 256 & -3.7 & 53 & 72 (6) & 7.12 (0.11) & 3.0 (0.3)\\
        38$_{(11,27)}$-38$_{(10,28)}$ & 234554.566 & 489 & -3.7 & 77 & 50 (4) & 6.72 (0.14) & 3.0 (-)\\
        25$_{(10,15)}$-25$_{(9,16)}$ & 234704.065 & 242 & -3.7 & 77 & 65 (11) & 6.5 (0.2) & 3.3 (0.6)\\
        34$_{(4,30)}$-34$_{(3,31)}$ & 246395.021 & 344 & -3.8 & 69 & 94 (10) & 7.98 (0.17) & 4.9 (0.5)\\
        34$_{(5,30)}$-34$_{(4,31)}$ & 246605.634 & 344 & -3.8 & 69 & 90 (7) & 7.32 (0.15) & 4.4 (0.4)\\
        30$_{(2,28)}$-30$_{(1,29)}$ & 246773.215 & 253 & -4.0 & 61 &  70 (4) & 7.01 (0.09) & 3.7 (0.2)\\
        30$_{(3,28)}$-30$_{(2,29)}$ & 246778.410 & 253 & -4.0 & 61 & 60 (6) & 7.27 (0.16) & 3.9 (0.4)\\
        36$_{(11,26)}$-36$_{(10,27)}$ & 246852.728 & 446 & -3.6 & 73 & 81 (7) & 6.83 (0.10) & 3.3 (0.2)\\
        32$_{(3,29)}$-32$_{(2,30)}$ & 247285.434 & 297 & -3.9 & 65 & 92 (9) & 6.64 (-) & 3.5 (-)\\
        32$_{(4,29)}$-32$_{(3,30)}$ & 247323.158 & 297 & -3.9 & 65 & 50 (11) & 6.73 (-) & 3.5 (-)\\
        \noalign{\vskip 0.5mm}
        \specialrule{.2em}{.1em}{.1em}
\end{longtable}
\begin{tablenotes}
\footnotesize
\item $^{a}$Spectroscopic parameters of CH$_3$OH \citep{methanol_ref} and CH$_2$OHCHO \citep{glyco_ref1, glyco_ref2} are retrieved from the CDMS molecular database \citep[Cologne Database for Molecular Spectroscopy]{cdms_doc}, and the ones of HCOOCH$_3$ \citep{methylformate_ref}, CH$_3$CHO \citep{acetaldehyde_ref}, NH$_2$CHO \citep{formamide_ref1, formamide_ref2, formamide_ref3, formamide_ref4, formamide_ref5}, and CH$_3$CH$_2$OH \citep{ethanol_ref} from the JPL molecular database \citep[Jet Propulsion Laboratory]{pickett_1998, jpl_doc}. $^{b}$Results of the Gaussian fit algorithm in \texttt{CARTA}. Note that the spectral resolution is 1.4 km s$^{-1}$. In round parentheses are reported the uncertainties on the fitted quantities. The dash symbol indicates that the quantity has been fixed during the line fitting procedure. $^{c}$ Refers to the transition detected in the narrow SPW centred at 260 GHz.
\end{tablenotes}

\section{Image plane fit results}\label{appendixB}

\begin{longtable}{lcccccc}
\caption{Image plane Gaussian fitted parameters for selected transitions.}\label{tab3}  \\
\specialrule{.2em}{.1em}{.1em}
Transition & E$_{\rm u}$ & R.A. & Dec. & $\theta_{M}$ & $\theta_{m}$ & P.A.\\
 & [K] & [h: m: s] & [$^{\circ}:^{'}:^{"}$] & [mas] & [mas] & [$^{\circ}$]\\
\noalign{\vskip 1mm}
\hline
\endfirsthead
\caption{continued.}\\
\endhead 
\multicolumn{7}{c}{CH$_3$OH} \\
\hline
10$_{(3,7)}$-11$_{(2,9)}$ E  & 191 & 03:29:10.431 (0.002") & +31.13.32.023 (0.002") & 331 (7) & 224 (6) & 130 (3)\\
18$_{(3,15)}$-17$_{(4,14)}$ A  & 447 & 03:29:10.4306 (0.0013") & +31.13.32.0178 (0.0014") & 257 (6) & 207 (8) & 107 (6)\\
4$_{(2,3)}$-5$_{(1,4)}$ A & 61 & 03:29:10.431 (0.002") & +31.13.32.022 (0.002") & 351 (8) & 232 (7) & 129 (3)\\
5$_{(4,2)}$-6$_{(3,3)}$ E & 123 &  03:29:10.431 (0.002") & +31.13.32.023 (0.002") & 315 (8) & 229 (8) & 121 (4) \\
16$_{(-2,15)}$-15$_{(-3,13)}$ E  & 338 & 03:29:10.430 (0.002") & +31.13.32.0118 (0.0019") & 302 (7) & 224 (6) & 122 (4)\\
4$_{(2,2)}$-5$_{(1,5)}$ A & 61 & 03:29:10.430 (0.003") & +31.13.32.019 (0.003") & 353 (9) & 239 (6) & 130 (3)\\
18$_{(3,15)}$-18$_{(2,16)}$ A  & 447 & 03:29:10.430 (0.002") & +31.13.32.017 (0.002") & 310 (8) & 224 (6) & 125 (4)\\
\noalign{\vskip 0.5mm}
\hline 
\noalign{\vskip 0.5mm}
\multicolumn{7}{c}{HCOOCH$_3$} \\
\noalign{\vskip 0.5mm}
\hline 
19$_{4,16}$-18$_{4,15}$ A  & 123 & 03:29:10.432 (0.002") & +31.13.32.010 (0.002") & 322 (8) & 231 (8) & 118 (4)\\
18$_{4,14}$-17$_{4,13}$ E  & 114 & 03:29:10.432 (0.002") & +31.13.32.010 (0.002") & 313 (8) & 229 (8) & 117 (4)\\
19$_{(11,9)}$-18$_{(11,8)}$ E  & 193 & 03:29:10.432 (0.003") & +31.13.32.010 (0.003") & 277 (11) & 220 (13) & 105 (10)\\
20$_{(13,8)}$-19$_{(13,7)}$ E  & 236 & 03:29:10.431 (0.003") & +31.13.32.004 (0.003") & 258 (13) & 204 (13) & 107 (11)\\
20$_{(6,15)}$-19$_{(6,14)}$ A [v=1] & 336 &  03:29:10.431 (0.004") & +31.13.32.004 (0.004") & 247 (16) & 204 (16) & 106 (17)\\
\noalign{\vskip 0.5mm}
\hline 
\noalign{\vskip 0.5mm}
\multicolumn{7}{c}{CH$_3$CHO} \\
\noalign{\vskip 0.5mm}
\hline 
12$_{(2,11)}$-11$_{(2,10)}$ E [$v_t$=1]  & 285 & 03:29:10.4311 (0.0011") & +31.13.32.0064 (0.0014") & 206 (4) & 182 (6) & 91 (12)\\
12$_{(4,8)}$-11$_{(4,7)}$ E [$v_t$=2] & 487 & 03:29:10.4312 (0.0018") & +31.13.31.998 (0.002") & 186 (11) & 165 (13) & 52 (26)\\
13$_{(0,13)}$-12$_{(0,12)}$ E [$v_t$=2] & 461 & 03:29:10.4303 (0.0016") & +31.13.31.9957 (0.0018") & 160 (8) & 157 (9) & 73 (75)\\
14$_{(0,14)}$-13$_{(1,13)}$ A & 96 & 03:29:10.431 (0.003") & +31.13.31.999 (0.002") & 249 (10) & 192 (11) & 95 (9)\\
\noalign{\vskip 0.5mm}
\hline 
\noalign{\vskip 0.5mm}
\multicolumn{7}{c}{NH$_2$CHO} \\
\noalign{\vskip 0.5mm}
\hline
11$_{(4,7)}$-10$_{(4,6)}$  & 115 & 03:29:10.4306 (0.0008") & +31.13.32.0060 (0.0012") & 165 (6) & 153 (5) & 154 (19) \\
11$_{(3,9)}$-10$_{(3,8)}$  & 94 & 03:29:10.4308 (0.0008") & +31.13.32.0090 (0.0012") & 165 (6) & 139 (5) & 149 (9)\\
12$_{(0,12)}$-11$_{(0,11)}$  & 78 & 03:29:10.4303 (0.0010") & +31.13.32.0017 (0.0012") & 158 (5) & 136 (3) & 148 (10)\\
\noalign{\vskip 0.5mm}
\hline 
\noalign{\vskip 0.5mm}
\multicolumn{7}{c}{CH$_3$CH$_2$OH} \\
\noalign{\vskip 0.5mm}
\hline
g-20$_{(4,17)}$-20$_{(3,17)}$ [$v_t$=1-0] & 256 & 03:29:10.4308 (0.0012") & +31.13.32.0075 (0.0017") & 185 (8) & 165 (8) & 36 (16) \\
t-13$_{(5,9)}$-13$_{(4,10)}$ & 108 & 03:29:10.4312 (0.0014") & +31.13.32.0128 (0.0016") & 210 (6) & 179 (8) & 80 (11)\\
t-12$_{(5,8)}$-12$_{(4,9)}$ & 97 & 03:29:10.4311 (0.0011") & +31.13.32.0078 (0.0013") & 207 (5) & 175 (7) & 80 (9) \\
t-25$_{(5,21)}$-25$_{(4,22)}$ & 306 & 03:29:10.4310 (0.0009") & +31.13.32.0081 (0.0013") & 169 (6) & 162 (6) & 44 (37)\\
g-14$_{(3,11)}$-13$_{(3,10)}$ [$v_t$=0-0] & 156 & 03:29:10.4302 (0.0018") & +31.13.32.0064 (0.0018") & 200 (9) & 169 (10) & 83 (11)\\
\noalign{\vskip 0.5mm}
\hline
\noalign{\vskip 0.5mm}
\multicolumn{7}{c}{CH$_2$OHCHO} \\
\noalign{\vskip 0.5mm}
\hline
26$_{(10,17)}$-26$_{(9,18)}$  & 256 & 03:29:10.4315 (0.0013") & +31.13.32.001 (0.002") & 154 (10) & 144 (8) & 16 (79)\\
38$_{(11,27)}$-38$_{(10,28)}$  & 489 & 03:29:10.4318 (0.0017") & +31.13.32.004 (0.003") & 169 (12) & 155 (4) & 13 (89)\\
34$_{(5,30)}$-34$_{(4,31)}$  & 344 & 03:29:10.4304 (0.0019") & +31.13.32.001 (0.002") & 153 (9) & 139 (7) & 158 (51)\\
30$_{(2,28)}$-30$_{(1,29)}$  & 253 & 03:29:10.431 (0.002") & +31.13.31.995 (0.003") & 149 (14) & 129 (17) & 30 (37)\\
36$_{(11,26)}$-36$_{(10,27)}$ & 446 & 03:29:10.4308 (0.0018") & +31.13.31.997 (0.003") & 143 (12) & 115 (13) & 16 (19)\\
\noalign{\vskip 0.5mm}
\specialrule{.2em}{.1em}{.1em}
\end{longtable}
\begin{tablenotes}
\footnotesize
\item Each species transitions are listed in order of increasing rest-frame frequency. Uncertainties are indicated in round parentheses.
\end{tablenotes}

\begin{figure*}[htp!]
    \centering
    \includegraphics[width=\textwidth]
{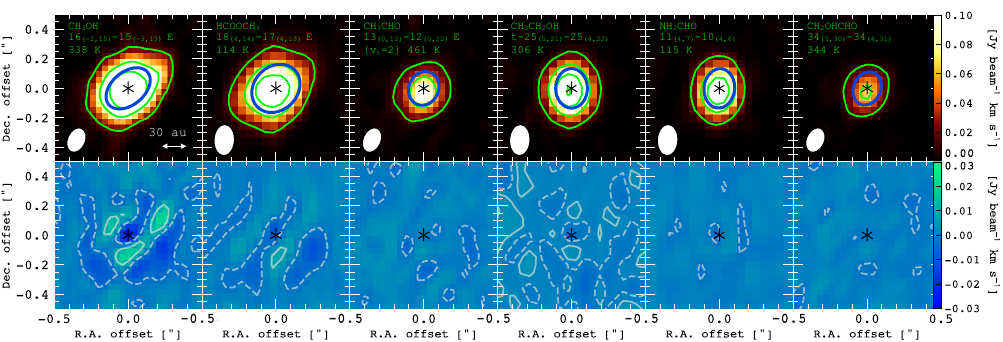}
    \caption{Examples of integrated intensity maps \textit{(left)} and image fit residuals \textit{(right)} in colour scale of the isolated iCOMs lines from which the emission size has been extracted. \textit{Left:} First contours and steps are respectively 5$\sigma$ and 20$\sigma$. The black marker corresponds to the centroid of the map. The blue solid line is the fitted Gaussian ellipse convolved with the beam. The synthesized beams refer to S1 (0$''$.21 $\times$ 0$''$.14, -3$^{\circ}$) and S2 (0$''$.17 $\times$ 0$''$.12, -28$^{\circ}$). \textit{Right:} Dashed and solid contours are respectively the -1$\sigma$ and 3$\sigma$ levels of the residual map. The black marker corresponds to the centroid of the map. }
    \label{fig:12}
\end{figure*}
\begin{figure*}[htp!]
    \centering
    \includegraphics[width=\textwidth]
{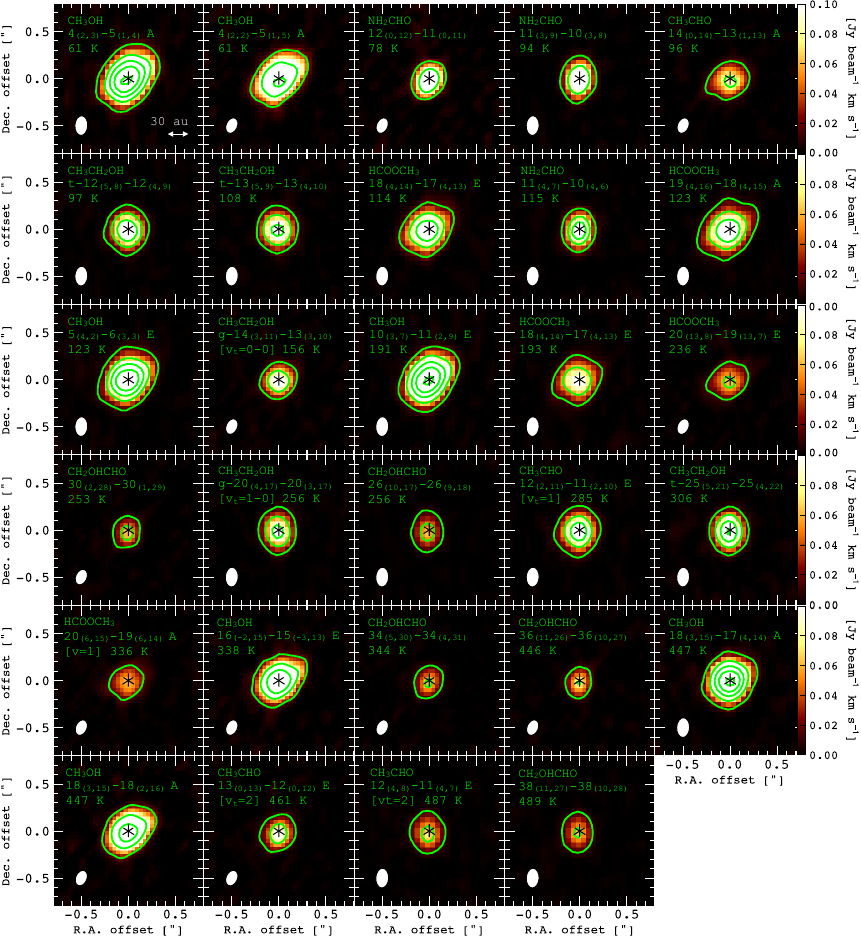}
    \caption{Integrated intensity maps in colour scale and green contours of the isolated iCOMs lines from which the emission size has been extracted. The maps are plotted in order of increasing upper-state energy. First contours and steps are respectively 5$\sigma$ and 20$\sigma$. The black marker corresponds to the centroid of the map. The synthesized beams refer to S1 (0$''$.21 $\times$ 0$''$.14, -3$^{\circ}$) and S2 (0$''$.17 $\times$ 0$''$.12, -28$^{\circ}$).}
    \label{fig:13}
\end{figure*}

\twocolumn
\section{Non-LTE LVG Analysis for CH$_3$OH}\label{appendixC}
Since methanol (CH$_3$OH) is known to be very abundant in this source \citep[see e.g., ][]{marta_2020} and optically-thick, to complement and check the analysis made with the LTE and optically-thin line assumptions, we performed a non-LTE analysis using a Large Velocity Gradient (LVG) code \citep{ceccarelli_2003}. We could therefore derive the physical properties of the gas emitting CH$_3$OH, namely gas temperature, density and column density, and the optical depth of the transitions. 
We used the collisional coefficients of both A-type and E-type of CH$_3$OH with para-H2, computed by \citet{rabli_rotational_2010} between 10 and 200 K for the first 256 levels and provided by the BASecOL database \citep{dubernet_2013}. Note that, given that the coefficient are available only for transition with J up to 15 and computed for temperatures up 200 K, we used the transitions that in this work have E$_{\rm u}$ less than 200 K. For this purpose we added the transition identified in the narrow SPW centered at 244 GHz, identified as 5$_{(1,3)}-4_{(1,3)}$ A with the following spectroscopic and Gaussian parameters: E$_{\rm u}$ = 50 K; Log$_{10}$A$_{\rm ul}$= -4.7; g$_{\rm u}$ = 44; $W$ = 340 $\pm$ 10 K km s$^{-1}$; V$_{peak}$ = 6.74 $\pm$ 0.08 km s$^{-1}$; FWHM = 5.8 $\pm$ 0.2 km s$^{-1}$. 
We assumed a spherical geometry to compute the line escape probability, the CH$_3$OH-A/CH$_3$OH-E ratio equal to 1, and the H$_2$ ortho-to-para ratio equal to 3.
We ran a large rid of models ($\sim$13,000) covering the frequency of the observed CH$_3$OH lines, a total (A-type plus E-type) column density $N_{\rm CH_3OH}$ from 10$^{16}$ to $4\times 10^{19}$ cm$^{-2}$, a gas density $n_{H_2}$ from 10$^{6}$ to $2\times 10^{8}$ cm$^{-3}$, both sampled in logarithmic scale, and a gas temperature T from 80 to 190 K, sampled in linear scale. 
We then simultaneously fitted the measured  CH$_3$OH-A and CH$_3$OH-E line intensities via comparison with those simulated by the LVG model, leaving $N_{\rm CH_3OH}$, $n_{\rm H_2}$, and T as free parameters.
Following the observations, we assumed a source size of $0\farcs3$ to compute the filling factor, a linewidth equal to 4.5 km s$^{-1}$, and we included the calibration/continuum subtraction uncertainty (20\%) in the observed intensities.

The chi-square best fit is obtained for a total CH$_3$OH column density of 2 $\times$ 10$^{18}$ cm$^{-2}$, a gas temperature of 100 K and gas density of 2 $\times$ 10$^6$ cm$^{-3}$, with reduced chi-square $\tilde{\chi}^2$ = 0.6.  
Finally, we corrected the intensities for the foreground dust absorption. In this case, the best fit is obtained for a total CH$_3$OH column density of 4 $\times$ 10$^{18}$ cm$^{-2}$,  a gas temperature of 150 K and gas density of 5 $\times$ 10$^6$ cm$^{-3}$, with reduced chi-square $\tilde{\chi}^2$ = 0.3. 
The results do not change assuming a linewidth $\pm$0.5 km s$^{-1}$ with respect to the chosen one. 
Figure \ref{fig:3} shows the density-temperature $\chi^2$ surface of the $N_{\rm CH_3OH}$ best fit. 
The fit results in the 1$\sigma$ confidence range are reported in Tab. \ref{tab1}. 

\section{LTE Population Diagram analysis}\label{appendixD}
We performed a population diagram analysis on methyl formate (HCOOCH$_3$), ethanol (CH$_3$CH$_2$OH), and glycolaldehyde (CH$_2$OHCHO) to correct rotational temperature T$_{\rm rot}$ and total column density N$_{\rm tot}$ for the line optical depth ($\tau$), following the prescription of \citet{goldsmith_1999}.  
The optical depth of a transition can be written as:
\begin{equation}\label{eq:1}
    \tau = \frac{c^3}{\rm 8 \pi \nu_0^{3}} \frac{A_{\rm ul}}{\rm \Delta V/(2 \sqrt{2~ln2})}   \frac{N_{\rm u}}{g_{\rm u}} \left( e^{\rm ~h \nu_0/kT_{\rm rot}} -1\right)
\end{equation}
where 
\begin{equation}\label{eq:nu}
    N_{\rm u}=W\times\frac{\rm 8\pi k \nu_0^2}{ h~c^3~A_{\rm ul}~{ff}} \times C_{\tau} \ .
\end{equation}
In the above, $c$ is the speed of light, $\nu_{\rm 0}$ the rest-frame line frequency, $A_{\rm ul}$ the Einstein coefficient for spontaneous emission, $\Delta$V the line profile FWHM (derived by fitting the line with a gaussian profile), g$_{\rm u}$ the statistical weight of the upper state, E$_{\rm u}$ the upper-state energy of the transition, $h$ and $k$ respectively the Planck and the Boltzmann constants, $W$ the velocity-integrated line intensity, and $ff$ the beam filling factor, defined as: $ff=\theta_{\rm source}^2/(\theta_{\rm source}^2+\theta_{\rm beam}^2)$ \citep[e.g.][]{mangum_2015}. $\theta_{\rm beam}$ is the synthesized beam of the observations. All units refer to the cgs system.

Finally, C$_{\tau}$ is the optical depth correction factor:
\begin{equation}\label{eq:2}
    C_{\tau} = \frac{1-e^{-\tau}}{\tau}
\end{equation}
This factor indicates how much the upper level populations (N$_{\rm u}$) are underestimated due to the line opacity. When
the line is optically-thin, C$_{\rm \tau}$ is equal to unity.

For a molecule in LTE, all excitation temperatures are the same, and the population of each level is given by:
\begin{equation}\label{eq:3}
    \ln\left( \frac{N_{\rm u}}{g_{\rm u}} \right) = \ln\left( \frac{N_{\rm tot}}{Q({T_{\rm rot}})} \right) - \frac{E_{\rm u}}{k T_{\rm rot}}
\end{equation}
where N$_{\rm tot}$ the species total column density, and Q(T$_{\rm rot}$) the partition function at the rotational temperature T$_{\rm rot}$ of the species.

\begin{figure*}[htp!]
    \centering
    \includegraphics[width=\textwidth]{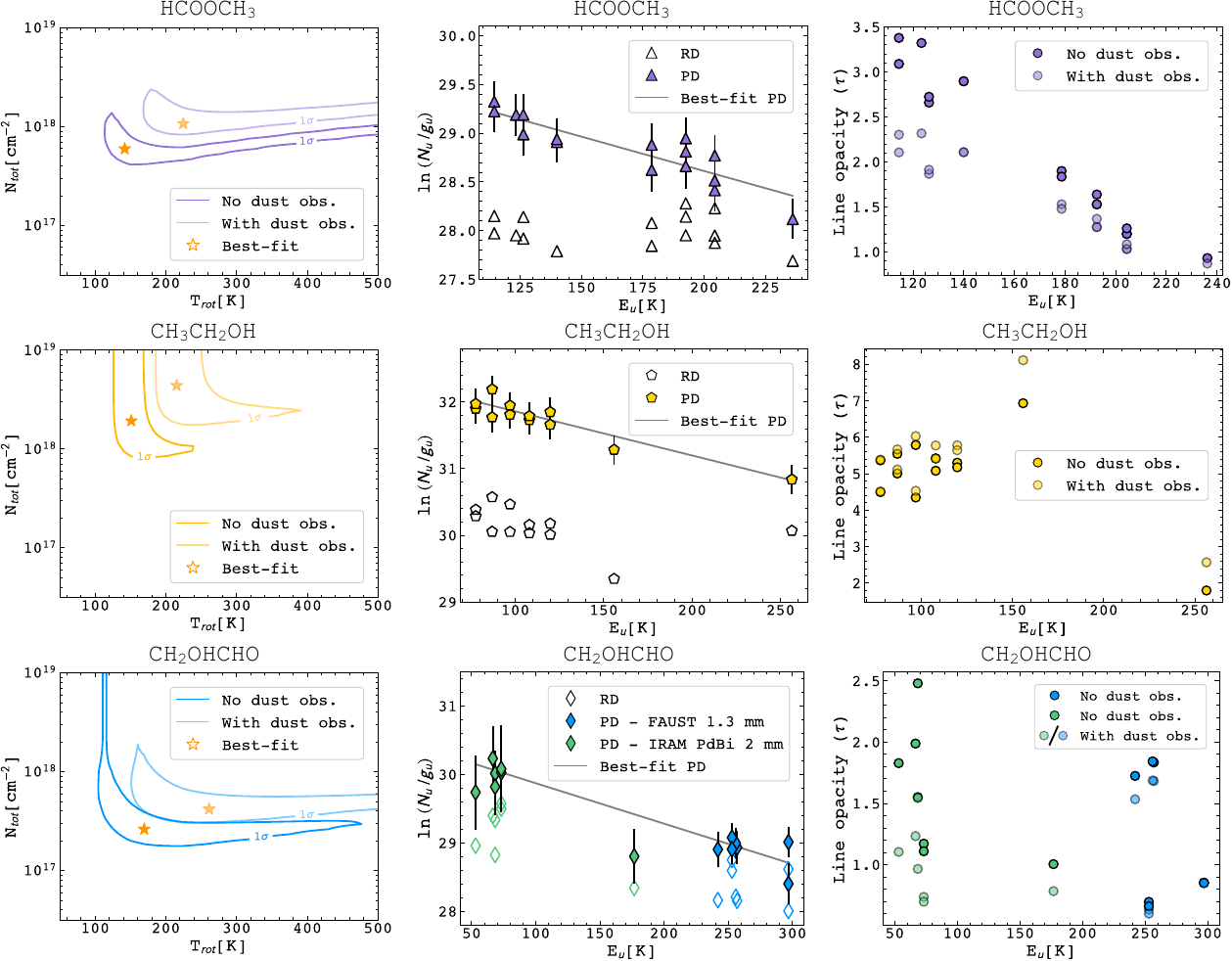}
    \caption{Same as Fig. \ref{fig:4} for lines with E$_{\rm u} <$ 300 K.}
    \label{fig:14}
\end{figure*}

We created a 2D parameter space grid in rotational temperature (50 values between 50 and 500 K), and in total column density (50 values between 1 $\times$ 10$^{16}$ and 5 $\times$ 10$^{19}$ cm$^{-2}$) at fixed source size $\theta_{\rm source}$ (see Tab. \ref{tab1}).
For each set of T$_{\rm rot}$ and N$_{\rm tot}$ we computed the model upper state column densities N$_{\rm u,~model}$ using equation \ref{eq:2}, as well as the line opacity $\tau$ (Eq. \ref{eq:1}) and the resulting C$_{\tau}$ (Eq. \ref{eq:2}). 
Then we retrieved the corrected upper state column densities N$_{\rm u,~corr}$ using Eq. \ref{eq:nu}, in which the observed velocity-integrate integrated line intensities are corrected for the beam dilution and the opacity factor.
Finally, we performed a chi-square minimization test comparing N$_{\rm u,~corr}$ and N$_{\rm u,~model}$:
\begin{equation}\label{eq:4}
    \tilde{\chi}^2 = \frac{1}{\rm N-3} \sum_{i=1}^{N} \frac{\rm ln(N_{u, corr, i}/g_{u, i}) - ln(N_{u, model, i}/g_{u, i})}{\rm \sigma_i^2}
\end{equation}
Here, $N$ is the number of observed lines, while $\sigma$ is the observed error on ln(N$_{\rm u}$/g$_{\rm u}$), calculated as:
\begin{equation}\label{eq:5}
    \sigma = \frac{\sqrt{(\Delta W)^2+\sigma_f^2}}{W}
\end{equation}
where $\Delta W$ is the error on the fitted line intensities (see Tab. \ref{tab2}), and $\sigma_f$ is the absolute flux calibration error, that we fixed as 20\% of $W$.
In Figs. \ref{fig:4} and \ref{fig:14} respectively, we report the results of this analysis accounting respectively for all transitions and the low excitation ones (E$_{\rm u}<$ 300 K) only. 

Accounting solely for lines with E$_{\rm u}\leq$ 300 K (see Tab. \ref{tab4}), T$_{\rm rot}$ and N$_{\rm tot}$ are strongly degenerate in the case of HCOOCH$_3$ and CH$_2$OHCHO, and less degenerate for CH$_3$CH$_2$OH. The resulting minimum $\tilde{\chi}^2$ is 0.6 (HCOOCH$_3$), 0.4 (CH$_3$CH$_2$OH), and 0.6 (CH$_2$OHCHO). We could extract only a lower limit on the temperature for HCOOCH$_3$, and a lower limit on the total column density for CH$_3$CH$_2$OH and CH$_2$OHCHO. 

The inclusion of high excitation lines (Fig. \ref{fig:4}) helps closing the $\tilde{\chi}^2$ contours within the investigated parameter space, and to attenuate the degeneracy between T$_{\rm rot}$, N$_{\rm tot}$ and $\tau$. The minimum $\tilde{\chi}^2$ in this case is 1.0 (HCOOCH$_3$), 0.9 (CH$_3$CH$_2$OH), and 0.5 (CH$_2$OHCHO). 

Accounting for a 50\% millimeter dust obscuration factor~\footnote{30\%, according to \citet{marta_2020}, for the sample of CH$_2$OHCHO lines from \citet{taquet_2015}.} on the velocity-integrated line intensities, the effect is to increase the best-fit temperature range by 50-100 K within error bars, and increase the total column density by a factor of about 1.6 (2.2 for CH$_3$CH$_2$OH). The best-fit $\tilde{\chi}^2$ is 0.6 (HCOOCH$_3$), 0.6 (CH$_3$CH$_2$OH), and 0.6 (CH$_2$OHCHO) for E$_{\rm u}$ $<$ 300 K, and 1.4 (HCOOCH$_3$), 1.3 (CH$_3$CH$_2$OH), and 0.5 (CH$_2$OHCHO) using all lines.
The effect on the line optical depth is to change the average line opacity by 0.4 dex (E$_{\rm u}$ $<$ 300 K) and 0.1 dex (all lines).

\begin{table}[htp!]
    \caption{1$\sigma$ confidence level results of the population diagram analysis with E$_{u} <$ 300 K.} \label{tab4}
    \centering
    \resizebox{0.9\columnwidth}{!}{%
    \begin{tabular}{lcccc}
    \specialrule{.2em}{.1em}{.1em}
    Species & \multicolumn{2}{c}{No dust obscuration} & \multicolumn{2}{c}{With dust obscuration}\\
    \cmidrule(r){2-3}
    \cmidrule(r){4-5}
    & T$_{\rm rot}$ & N$_{\rm tot}$ & T$_{\rm rot}$ & N$_{\rm tot}$ \\
    & [K] & [cm$^{-2}$] & [K] & [cm$^{-2}$]\\
    \noalign{\vskip 0.5mm}
    \hline
    \noalign{\vskip 0.5mm}
    HCOOCH$_3$ & $>$110 & 0.4-1.5 $\times$ 10$^{18}$ & $>$170 & 0.8-2.4 $\times$ 10$^{18}$\\
    CH$_3$CH$_2$OH & 130-240 & $>$8 $\times$ 10$^{17}$ & 190-400 & $>$1.8 $\times$ 10$^{18}$\\
    CH$_2$OHCHO & 110-480 & $>$2 $\times$ 10$^{17}$ & $>$160 & $>$3 $\times$ 10$^{17}$ \\
    \specialrule{.2em}{.1em}{.1em}
    \end{tabular}
    }
    \footnotesize
    \begin{tablenotes}
    \item The source size derived from image fitting is assumed for each species (see Sec. \ref{subsec:3.1}).  
    \end{tablenotes}
    \vspace{-0.5cm}
\end{table}

\clearpage
\newpage
\section{Additional notes and results}\label{appendixE}
\begin{figure}[htp!]
    \centering
    \includegraphics[width=0.33\textwidth]{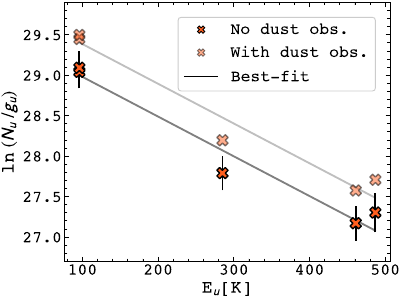}
    \caption{Rotational diagram (RD) of CH$_3$CHO. The solid gray line is the best-fit to the data points with \textit{(shaded)} and without \textit{(full color)} mm dust obscuration factor. The upper-state column densities are corrected for beam dilution.}
    \label{fig:15}
\end{figure}

Figure \ref{fig:15} shows the rotational diagram of CH$_3$CHO computed using all lines (see Tab. \ref{tab2}), and correcting for the derived source size of 0$''$.20. Shaded data points and solid line take into account the mm dust obscuration (see Sect. \ref{subsec:4.1}). The best-fit results in both cases are reported in Tab. \ref{tab1}. 

\begin{figure}[htp!]
    \centering
    \includegraphics[width=0.33\textwidth]{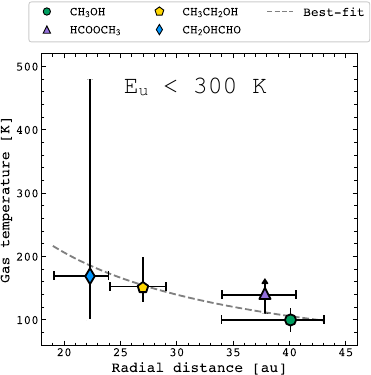}
    \caption{Same as Fig. \ref{fig:5} from iCOMs transitions with E$_{u} < 300$ K. The arrow indicates a lower limit on the temperature. The grey dotted line is the best-fit to the data points in the top panel of Fig. \ref{fig:5}).}
    \label{fig:16}
\end{figure}

Figure \ref{fig:16} shows the gas radial temperature profile of IRAS 4A2 obtained from the population diagram analysis of HCOOCH$_3$, CH$_3$CH$_2$OH, and CH$_2$OHCHO below 300 K, and the non-LTE LVG analysis of CH$_3$OH. CH$_3$CHO and NH$_2$CHO are here not reported, as it was not possible to perform a conservative line analysis or obtain an independent estimate of the gas temperature, respectively. The grey dotted line refers to the best-fit to the data points of the top panel in Fig. \ref{fig:5}.


\end{appendix}

\end{document}